\documentclass[aps]{revtex4-1}
\usepackage{graphicx}
\usepackage{natbib}

\newcommand{\lsim}{\stackrel{<}{_\sim}}
\newcommand{\gsim}{\stackrel{>}{_\sim}}
\def\beq{\begin{equation}}
\def\eeq{\end{equation}}
\def\bea{\begin{eqnarray}}
\def\eea{\end{eqnarray}}
\def\ve{\vert}
\def\vel{\left|}
\def\ver{\right|}
\def\nnb{\nonumber}
\def\ga{\left(}
\def\dr{\right)}
\def\aga{\left\{}
\def\adr{\right\}}
\def\rar{\rightarrow}
\def\nnb{\nonumber}
\def\la{\langle}
\def\ra{\rangle}
\def\ba{\begin{array}}
\def\ea{\end{array}}
\def\bea{\begin{eqnarray}}
\def\eea{\end{eqnarray}}
\def\tep{$b \rar s \ell^+ \ell^-$}
\def\ds{\displaystyle}
\def\ve{\vert}
\def\vel{\left|}
\def\ver{\right|}
\def\nnb{\nonumber}
\def\ga{\left(}
\def\dr{\right)}
\def\aga{\left\{}
\def\adr{\right\}}
\def\rar{\rightarrow}
\def\nnb{\nonumber}
\def\la{\langle}
\def\ra{\rangle}
\def\lla{\left<}
\def\rra{\right>}

\begin{document}
\bibliographystyle{plainnat}
\def\beq{\begin{equation}}
\def\eeq{\end{equation}}
\def\bea{\begin{eqnarray}}
\def\eea{\end{eqnarray}}
\def\ve{\vert}
\def\vel{\left|}
\def\ver{\right|}
\def\nnb{\nonumber}
\def\ga{\left(}
\def\dr{\right)}
\def\aga{\left\{}
\def\adr{\right\}}
\def\rar{\rightarrow}
\def\nnb{\nonumber}
\def\la{\langle}
\def\ra{\rangle}
\def\lla{\left<}
\def\rra{\right>}
\def\ba{\begin{array}}
\def\ea{\end{array}}
\def\BcDll{$B_c \rar D_s \ell^+ \ell^-$}
\def\BcDsll{$B_c \rar D_s^{*} \ell^+ \ell^-$}
\def\decay{$B_c \rar D_s (D_s^{*}) \ell^+ \ell^-$}
\def\tepm{$B_c \rar D_s^{*} \mu^+ \mu^-$}
\def\tept{$B_c \rar D_s^{*} \tau^+ \tau^-$}
\def\ds{\displaystyle}
\begin{titlepage}
\title{ {\Large {\bf
Study of $B_c\rar D_s^{*}\,  \ell^+ \ell^-$ in
Single Universal Extra Dimension} }\vspace{1cm} }

\author {\small \ U.~O.~Yilmaz}
\email[e-mail:]{uoyilmaz@karabuk.edu.tr}
\affiliation{\small Physics Department, Karabuk University, 78100 Karabuk, Turkey  }

\begin{abstract}
The rare semileptonic \BcDsll decay is studied in the scenario of the universal extra dimension model with a single extra dimension in which inverse of the compactification radius R is the only new parameter. The sensitivity of differential branching ratio, total branching ratio, polarization and forward-backward asymmetries of final state leptons, both for muon and tau, to the compactification parameter is presented. For some physical observables uncertainty on the form factors and resonance contributions have been considered in the calculations. Obtained results, compared with the available data, show that there appear new contributions due to the extra dimension.
\end{abstract}

\maketitle \thispagestyle{empty}
\end{titlepage}

\section{Introduction}
Flavor-changing neutral current (FCNC) $b \rar s,d$ transitions which occur at loop level in the standard model (SM) provide us a powerful tool to test the SM and also a frame to study physics beyond the SM. After the observation of $b \rar s \, \gamma$ \cite{CLEO95}, these transitions became more attractive and since then rare radiative, leptonic and semileptonic decays of $B_{u,d,s}$ mesons have been intensively studied \cite{AAli05}. Among these decays, semileptonic decay channels are significant because of having relatively larger branching ratio.
The experimental data for exclusive $B\rar K^{(*)} \ell^+ \ell^-$ also increased the interest in these decays. These studies will be even more complete if similar studies for $B_c$, discovered by CDF Collaboration \cite{CDF98}, are also included.

The $B_c$ meson is the lowest bound state of two heavy quarks, bottom $b$ and charm $c$, with
explicit flavor that can be compared with the $c\bar{c}$ and $b\bar{b}$- bound state
which have implicit flavor. The implicit-flavor states decay
strongly and electromagnetically whereas the $B_c$ meson decays
weakly. $B_{u,d,s}$ are
described very well in the framework of the heavy quark limit,
which gives some relations between the form factors of the
physical process. In case of $B_c$ meson, the heavy flavor and
spin symmetries must be reconsidered because of heavy $b$ and $c$.
On the experimental side of the decay, for example, at LHC,  $10^{10} B_c $ events per year is estimated \cite{Sun}-\cite{Altarelli08}. This reasonable number is stimulating the work on the $B_c$ phenomenology and  this possibility will provide
information on rare $B_c$ decays as well as CP violation and polarization asymmetries.

In rare $B$ meson decays, effects of the new
physics may appear in two different manners, either through the
new contributions to the Wilson coefficients existing in the SM or
through the new structures in the effective Hamiltonian which are
absent in the SM.

Considering different models beyond the SM, extra dimensions are specially attractive because of including gravity and other interactions, giving hints on the hierarchy problem and a connection with string theory. Those with universal extra dimensions (UED) have special interest because all the SM particles propagate in extra dimensions, the compactification of which allows Kaluza-Klein (KK) partners of the SM fields in the four-dimensional theory and also KK modes without corresponding the SM partners \cite{Antoniadis90, Antoniadis98, Hamed98, Hamed99}. Throughout the UED, a simpler scenario with a single universal extra dimension is the Appelquist-Cheng-Dobrescu (ACD) model \cite{ACD}. The only additional free parameter with respect to the SM is the inverse of the compactification radius, $1/R$. In particle spectrum of the ACD model, there are infinite towers of KK modes and the ordinary SM particles are presented in the zero mode.

This only parameter have been attempted to put a theoretical or experimental restriction on it. Tevatron experiments put the bound $1/R \geq 300 GeV$. Analysis of the anomalous magnetic moment and $B\rar X_s \gamma$ \cite{Agashe} also lead to the bound $1/R \geq 300 GeV$. In the study of $B\rar K^* \gamma$ decay \cite{Colangelo06}, the results restrict R to be $1/R\geq250 GeV$. Also, in \cite{Haisch07} this bound is $1/R\geq 330 \,GeV$.
In two recent works, the theoretical study of $B \rar K \eta \gamma$ matches with experimental data if $1/R\geq 250 \,GeV$ \cite{Colangelo12} and  using the experimental result \cite{CDF} and theoretical prediction on the branching ratio of $\Lambda_b \rar \Lambda \mu^+ \mu^-$, the lower bound was obtained to be approximately $1/R\sim 250\, GeV$ \cite{Azizi12}. In this work, we will consider $1/R$ from 200 GeV up to 1000 GeV, however, under above consideration $1/R=250-350 \,GeV$ region will be taken more common bound region.
In literature, effective Hamiltonian of several FCNC processes \cite{Buras03, Buras04},
semileptonic and radiative decays have been investigated in the ACD model \cite{Colangelo06-2, Devidze06, Aliev07, Mohanta07, Colangelo08, Saddique08, Aslam08, Bashiry09, Azizi11, Azizi11-2, Li11}.

Concentrating on \BcDsll decay, it has been studied by using model independent effective Hamiltonian \cite{uoyilmaz07}, in Supersymmetric models \cite{Aslam11-08} and with fourth generation effects \cite{Aslam11-07}.
Also in \cite{Aslam11-01}, the UED effects on branching ratio and helicity fractions of the final state $D^*$ meson were calculated using the form factors obtained through the Ward identities for this process. The weak annihilation contribution in addition to the FCNC transitions was taken into account. We will, however, only consider the FCNC transitions and calculate the lepton asymmetries adding the resonance contributions.

The main aim of this paper is to find the effects of the ACD model on some physical observables related to the \BcDsll decay, while doing this we also give the behavior of these observables by a couple of figures in the SM.
Measurement of final state lepton polarizations is an useful way in searching new physics beyond the SM.
Another tool is the study of forward-backward asymmetry $(A_{FB})$, especially the position of zero value of $A_{FB}$ is very sensitive to the new physics. In addition to differential decay rate and branching ratio, we study forward-backward asymmetry and polarization of final state leptons, including resonance contributions and uncertainty on form factors in as many as possible cases.
We analyze these observables in terms of the compactification factor and the form factors.
The form factors for \BcDsll have been calculated using the light front,
constitute quark models \cite{Geng2002}, the relativistic constituent quark model \cite {Faessler}, relativistic quark model \cite{Ebert} and light-cone quark model \cite{Wang11}. In this work, we will use the form factors calculated in three-point QCD sum rules \cite{Azizi08}.

The paper is organized as follows. In Sec. II, we give the
effective Hamiltonian for the quark level process \tep and mention briefly the Wilson coefficients in the ACD model; a detailed discussion is given in Appendix A.
We drive matrix element using the form factors and calculate the decay rate in Sec. III.
In Sec. IV, we present the forward-backward asymmetry and Sec. V is devoted to lepton polarizations.
In the last section, we introduce our conclusions.

\section{Effective Hamiltonian and Wilson Coefficients}

The quark-level transition of  \BcDsll decay is governed by $b\rar s \ell^+ \ell^- $ and given
by the following effective Hamiltonian in the SM \cite{Buras96}
\bea
\label{effH} {\cal H}_{eff} &=&
            \frac{G_F \alpha}{\sqrt{2} \pi} V_{tb}V_{ts}^\ast
    \Bigg[ C_{9}^{eff} (\bar s \gamma_{\mu} L\, b)\, \bar \ell \gamma^\mu
      + C_{10} (\bar s \gamma_{\mu} L\, b)\, \bar \ell \gamma^\mu \gamma_5 \ell \nnb \\
      & & -2C_{7}^{eff} m_b (\bar s i \sigma_{\mu \nu} \frac {q^{\nu}} {q^2} R\, b)\,\bar \ell \gamma^\mu \ell
     \Bigg] \,,
\eea
where q is the momentum transfer, $L, R = (1 \pm \gamma_5)/2$ and ${C_i}$s are the
Wilson coefficients evaluated at the b quark mass scale.

The coefficient $C_9^{eff}$ has perturbative and resonance contributions.
So, $C_9^{eff}$ can be written as
\bea
C_9^{eff}(\mu) = C_9(\mu) \Big( 1+\frac{\alpha_{s}(\mu)} {\pi}\omega (s') \Big)+ Y(\mu, s')+C_9^{res}(\mu,s')
\eea
where $s'=q^2/m_{b}^2$.
\\
The perturbative part, coming from one-loop matrix elements of the four-quark operators, is
\bea
\label{EqY} Y(\mu,s')&=& h(y,s') [ 3 C_1(\mu) + C_2(\mu) + 3
       C_3(\mu) + C_4(\mu) + 3 C_5(\mu) + C_6(\mu)] \nnb \\
&-&     \frac{1}{2} h(1,s') \left( 4 C_3(\mu) + 4 C_4(\mu)
        + 3 C_5(\mu) + C_6(\mu) \right)\nnb \\
&- &  \frac{1}{2} h(0,s') \left[ C_3(\mu) + 3 C_4(\mu) \right] \nnb \\
&+&   \frac{2}{9} \left( 3 C_3(\mu) + C_4(\mu) + 3 C_5(\mu) +
      C_6(\mu) \right),
\eea
with $y=m_c/m_b$. The explicit forms of the functions $\omega(s')$ and $h(y,s')$ are given in \cite {Buras95}-\cite{Misiak93}. \\

The resonance contribution due to the conversion of the real $c \bar c$ into lepton pair can be done by using a Breit-Wigner shape as \cite{AAli91},
\begin{eqnarray}
C^{res}_{9}(\mu,s')&=&-\frac{3}{\alpha^2_{em}}\kappa \sum_{V_i=\psi_i}
    \frac{\pi \Gamma(V_i\rightarrow \ell^+ \ell^-)m_{V_i}}{s m_{b}^2 -m^{2}_{V_i}+i m_{V_i}
    \Gamma_{V_i}} \nonumber \\
&\times & [ 3 C_1(\mu) + C_2(\mu) + 3 C_3(\mu) + C_4(\mu) + 3
    C_5(\mu) + C_6(\mu)]\, .
 \label{Yresx}
\end{eqnarray}
The normalization is fixed by the data in \cite {pdg} and the phenomenological parameter $\kappa$ is taken 2.3 to produce the correct branching ratio $BR (B \rar J/\psi
K^* \rar K^* \ell^+ \ell^-)=BR (B \rar J/\psi K^* )B(J/\psi \rar \ell^+ \ell^-)$.

In the ACD model, there are not any new operators, therefore, new physics contributions appear by modifying the Wilson coefficients available in the SM. In this model, the Wilson coefficients can be written in terms of some periodic functions, as a function of compactification factor $1/R$. The function $F(x_t, 1/R)$ which generalize the $F_0(x_t)$ SM functions according to
\bea
\label{WACD} F(x_t, 1/R) = F_0(x_t) + \sum_{n=1} ^{\infty} F_n
(x_t, x_n)
\eea
where $x_t=m_t^2/m_W^2$, $x_n=m_n^2/m_W^2$ with the mass of KK particles $m_n=n/R$. n=0 corresponding the ordinary SM particles. The modified Wilson coefficients in the ACD model, taken place in many works in literature, are discussed in Appendix A.

Briefly, for $C_9$, in the ACD model and in the NDR scheme we have
\bea
\label{C9mu} C_9(\mu,1/R)= P_0^{NDR} + \frac {Y(x_t, 1/R)} {sin^2{\theta_W}} - 4 Z(x_t, 1/R) + P_E E(x_t, 1/R).
\eea
Instead of $C_7$, a normalization scheme independent effective coefficient $C_7^{eff}$ can be written as
\bea
\label{C7eff} C_7^{eff}(\mu,1/R) = &&\eta^{16/23} C_7(\mu_W, 1/R) \nnb \\
 &&+ \frac{8}{3} (\eta^{14/23} - \eta^{16/23}) C_8(\mu_W, 1/R) +
                    C_2(\mu_W, 1/R) \sum_{i=1} ^{8} h_i \eta^{a_i}.
\eea
The Wilson coefficient $C_{10}$ is independent of scale $\mu$ and given by
\bea
\label{C10} C_{10}(1/R) = - \frac{Y(x_t, 1/R)} {sin^2{\theta_W}}.
\eea
\begin{figure}[h]
\centering
\includegraphics[scale=0.62]{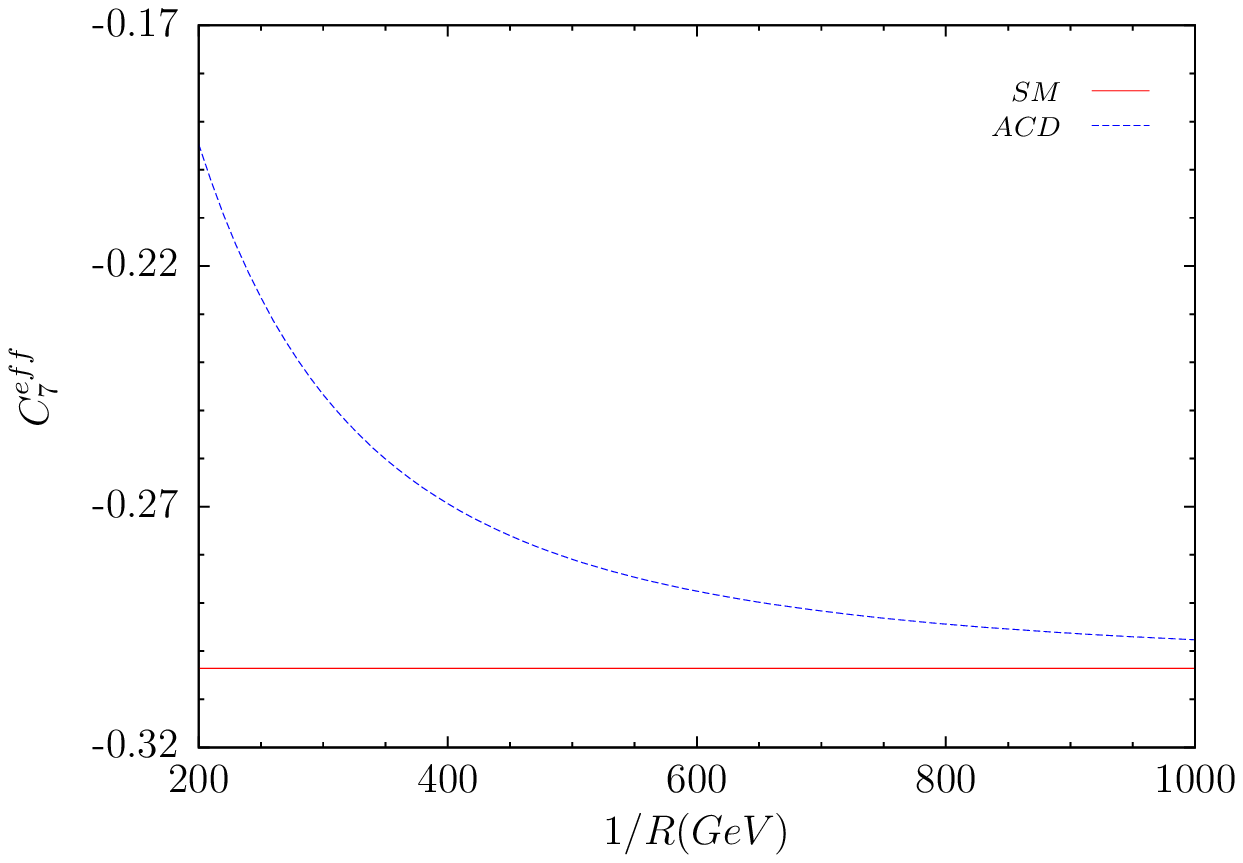}~~~~~
\centering
\includegraphics[scale=0.62]{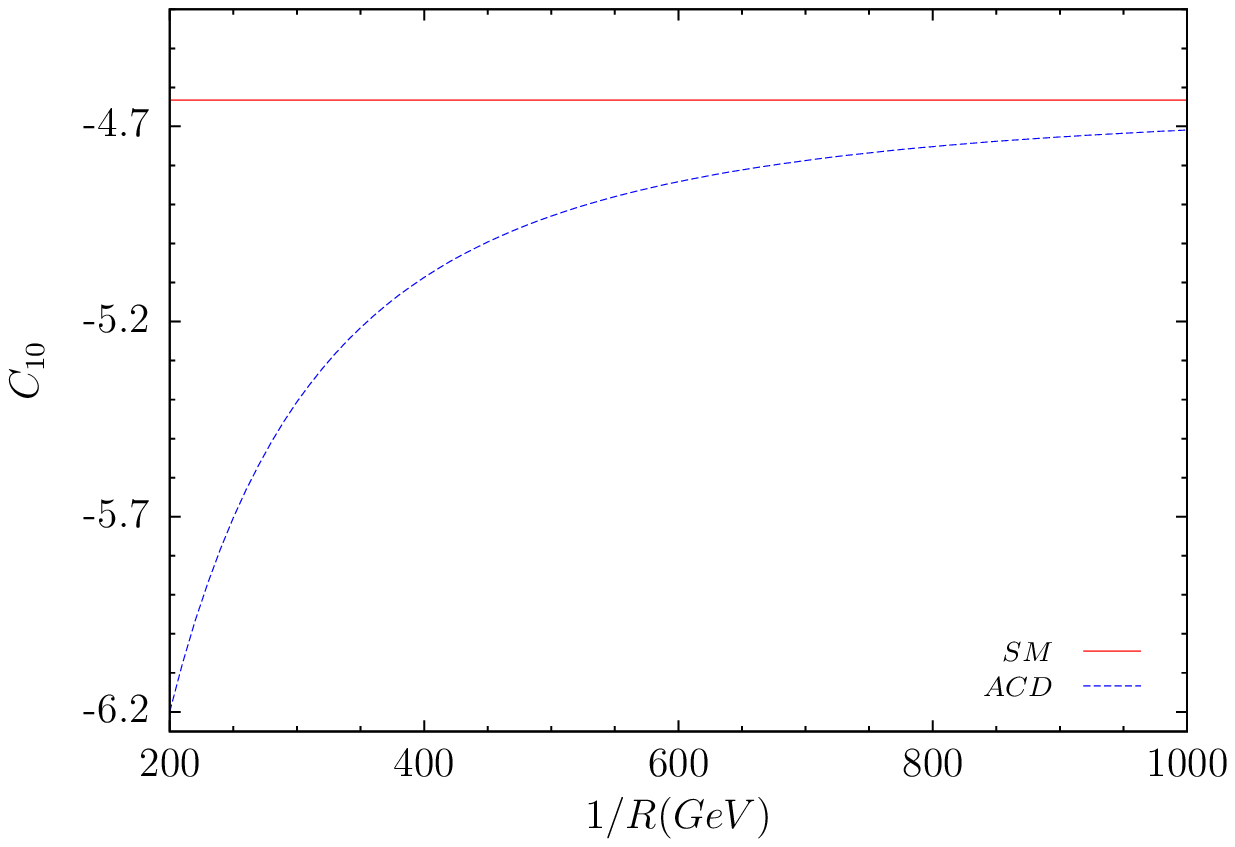}
\centering
\includegraphics[scale=0.62]{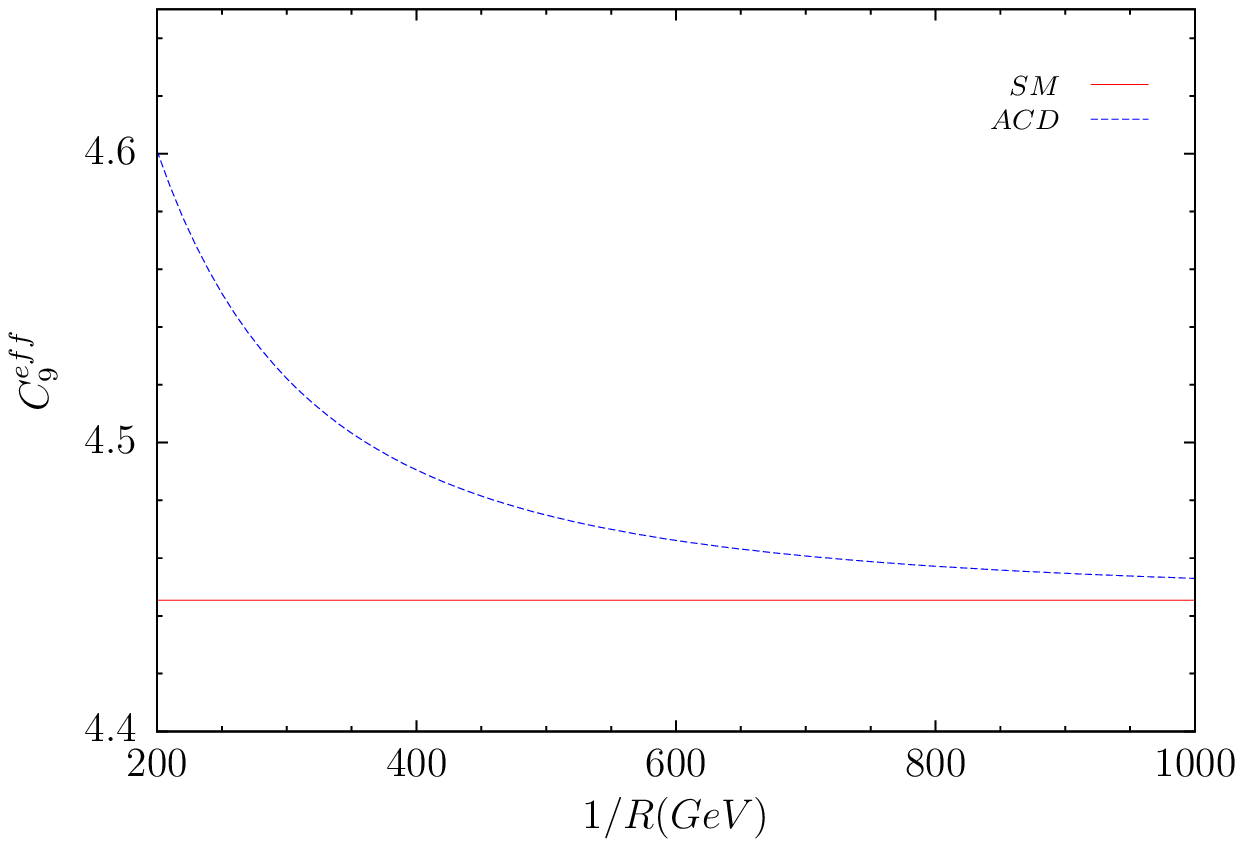}
\caption{The variation of Wilson coefficients with respect to $1/R$ at $q^2=14 \, GeV^2$
for the normalization scale $\mu=4.8 \,GeV$. ($C_9^{eff}$ does not include resonance contributions.)\label{coef}}
\end{figure}

The Wilson coefficients differ considerably from the SM values for small R.
The variation of modified Wilson coefficients with respect to $1/R$ at $q^2=14\,GeV^2$,
in which the normalization scale is fixed to $\mu=\mu_b\simeq 4.8 \,GeV$, is given in Fig. \ref{coef}.
The suppression of $\vel {C^{eff}_7}\ver$ for $1/R=250-350 \, GeV$ amount to $75\%-86\%$ relative to the SM value.
$\vel {C_{10}}\ver$ is enhanced by $23\%-13\%$. The impact of the ACD on $\vel {C^{eff}_9}\ver$ is very small.
For $1/R\gsim 600\, GeV$ the difference is less than $5\%$.
\section{Matrix Elements and Decay Rate}
The hadronic matrix elements in the exclusive \BcDsll decay can be obtained by
sandwiching the quark level operators in the effective Hamiltonian between the
initial and the final state mesons.
The nonvanishing matrix elements are parameterized in terms of form factors
as follows \cite{AAli2000, Ball05}
\bea
\label{bdsbir} \lla
    D_s^\ast(p_{D_s^\ast},\varepsilon) \vel \bar s \gamma_\mu
    (1 - \gamma_5) b \ver B_c(p_{B_c}) \rra =
- \epsilon_{\mu\nu\alpha\beta} \varepsilon^{\ast\nu}
    p_{D_s^\ast}^\alpha q^\beta \frac{2 V(q^2)}{m_{B_c}+m_{D_s^\ast}}
 - i \varepsilon_\mu^\ast (m_{B_c}+m_{D_s^\ast})
    A_1(q^2) \nnb \\
+ i (p_{B_c} + p_{D_s^\ast})_\mu (\varepsilon^\ast q)
    \frac{A_2(q^2)}{m_{B_c}+m_{D_s^\ast}}
 + i q_\mu (\varepsilon^\ast q)\frac{2 m_{D_s^\ast}}{q^2} [A_3(q^2) -
A_0(q^2)] ,
    \eea
and
\bea
\label{bdsiki} \lla D_s^\ast(p_{D_s^\ast},\varepsilon) \vel
    \bar s i \sigma_{\mu\nu} q^\nu
    (1 + \gamma_5) b \ver B_c(p_{B_c}) \rra =
2 \epsilon_{\mu\nu\alpha\beta} \varepsilon^{\ast\nu}
    p_{D_s^\ast}^\alpha q^\beta T_1(q^2) \nnb \\
+ i \Big[
    \varepsilon_\mu^\ast (m_{B_c}^2-m_{D_s^\ast}^2) -
    (p_{B_c} + p_{D_s^\ast})_\mu (\varepsilon^\ast q) \Big] T_2(q^2)
+ i (\varepsilon^\ast q) \Big[ q_\mu - (p_{B_c} +
    p_{D_s^\ast})_\mu \frac{q^2}{m_{B_c}^2-m_{D_s^\ast}^2} \Big] T_3 (q^2),
    \eea
where $q = p_{B_c}-p_{D_{s}^\ast}$ is the momentum transfer and
$\varepsilon$ is the polarization vector of $D_{s}^\ast$ meson.

The relation between the form factors $A_1(q^2)$, $A_2(q^2)$ and $A_3(q^2)$ can be stated as
\bea
\label{AA3}  A_3(q^2) = \frac {m_{B_s} + m_{\phi}}
                        {2 m_{\phi}}  A_1(q^2)
                - \frac {m_{B_s} - m_{\phi}} {2 m_{\phi}} A_2(q^2) \nnb
\eea
and in order to avoid kinematical singularity in the matrix element at
$q^2=0$, it is assumed that $A_0(0) = A_3(0)$ and $T_1(0)=T_2(0)$ \cite{Ball05}.\\

Using the effective Hamiltonian and matrix elements in Eqs. (\ref{bdsbir})--(\ref{bdsiki}), the transition amplitude for \BcDsll  is written as
 \bea
 \lefteqn{ \label{had}
    {\cal M}($\BcDsll$) =
    \frac{G \alpha}{2 \sqrt{2} \pi} V_{tb} V_{ts}^\ast }\nnb \\
&&\times \Bigg\{
    \bar \ell \gamma^\mu \ell \, \Big[
    -2 A \epsilon_{\mu\nu\alpha\beta} \varepsilon^{\ast\nu}
    p_{D_s^\ast}^\alpha q^\beta
    -i B \varepsilon_\mu^\ast
    + i C (\varepsilon^\ast q) (p_{B_c}+p_{D_s^\ast})_\mu
    + i D (\varepsilon^\ast q) q_\mu  \Big] \nnb \\
&&+ \bar \ell \gamma^\mu \gamma_5 \ell \, \Big[
    -2 E \epsilon_{\mu\nu\alpha\beta} \varepsilon^{\ast\nu}
    p_{D_s^\ast}^\alpha q^\beta
    -i F \varepsilon_\mu^\ast
    + i G (\varepsilon^\ast q) (p_{B_c}+p_{D_s^\ast})_\mu
    + i H (\varepsilon^\ast q) q_\mu  \Big]
    \Bigg\},
\eea
with the auxiliary functions
\bea \label{as} A &=& C_9^{eff}
    \frac{V(q^2)}{m_{B_c}+m_{D_s^\ast}} +
     \frac{2m_{b}}{q^2}C_7^{eff} T_1(q^2), \nnb \\
B &=& C_9^{eff}(m_{B_c}+ m_{D_s^\ast}) A_1(q^2)+
    \frac{2 m_{b}}{q^2}C_7^{eff} (m_{B_c}^2-m_{D_s^\ast}^2) T_2(q^2), \nnb \\
C &=& C_9^{eff} \frac {A_2(q^2)} {m_{B_c}+m_{D_s^\ast}} +
    \frac{2m_{b}}{q^2}C_7^{eff} \Big(T_2(q^2) + \frac {q^2}{m_{B_c}^2 - m_{D_s^\ast}^2} T_3(q^2) \Big),
    \nnb \\
D &=&  2 C_9^{eff} \frac{m_{D_s^\ast}} {q^2} (A_3(q^2) -A_0(q^2))-
     2 \frac{m_{b}} {q^2} C_7^{eff} T_3(q^2), \nnb
    \\
E &=& C_{10} \frac{V(q^2)}{m_{B_c}+m_{D_s^\ast}}, \nnb \\
F &=& C_{10}(m_{B_c}+m_{D_s^\ast}) A_1(q^2), \nnb \\
G &=& C_{10} \frac {A_2(q^2)} {m_{B_c}+m_{D_s^\ast}} , \nnb \\
H &=& 2 C_{10} \frac{m_{D_s^\ast}} {q^2} (A_3(q^2) -A_0(q^2)) .
\eea
Integrating over the angular dependence of the double differential decay rate, following dilepton mass spectrum is obtained
\bea
\label{bdsunp}
\frac{d
    \Gamma}{ds} = \frac{G^2 \alpha^2
        m_{B_c}}{2^{12} \pi^5 }
         \vel V_{tb} V_{ts}^\ast \ver^2 \sqrt{\lambda} v \Delta_{D_{s}^\ast}
\eea
where $s=q^2/m_{B_c}^2$, $\lambda= 1 + r^2 + s^2 -2r-2s-2rs$,
$r=m_{D_s^\ast}^2/m_{B_c}^2$, $v=\sqrt{1-{4m_\ell^2}/{s m_{B_c}^2}}$ and
\bea
\label{bdsdelta} \Delta_{D_{s}^\ast} =& & \frac{8} {3} \lambda m_{B_c}^6 s
\Big[(3-v^2) \vel A \ver^2 + 2v^2 \vel E \ver^2 \Big] + \frac{1} {r}
\lambda m_{B_c}^4
   \Bigg[\frac{1}{3} \lambda m_{B_c}^2 (3-v^2) \vel C \ver^2
        +m_{B_c}^2 s^2 (1-v^2) \vel H \ver^2 \nnb \\
  && +\frac {2}{3} \Big[(3-v^2)(r+s-1)-3s(1-v^2) \Big]Re[FG^\ast]
   +2m_{B_c}^2 s (1-r) (1-v^2) Re[GH^\ast]\nnb \\
                 && -2s(1-v^2) Re[FH^\ast]
      +\frac{2}{3}(3-v^2)(r+s-1)Re[BC^\ast] \Bigg]
         +\frac{1}{3r}(3-v^2) m_{B_c}^2 \Bigg[(\lambda +12r s)  \vel B \ver^2 \nnb \\
         &&+\lambda m_{B_c}^4 \Big[ \lambda -3s(s-2r-2) (1-v^2) \Big] \vel G \ver^2
                +\Big[ \lambda + 24r s v^2 \Big]\vel F \ver^2 \Bigg].
\eea
\begin{figure}[h]
\centering
\includegraphics[scale=0.62]{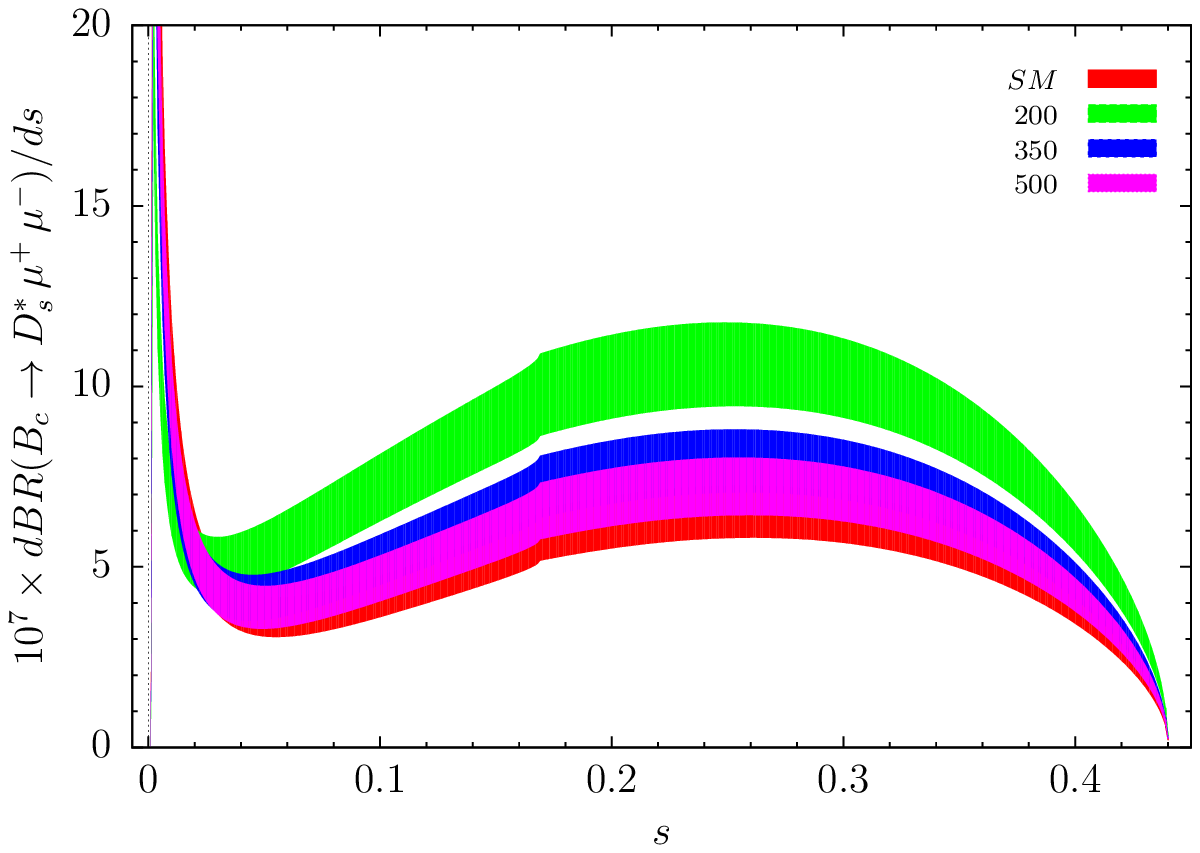}~~~~~
\includegraphics[scale=0.62]{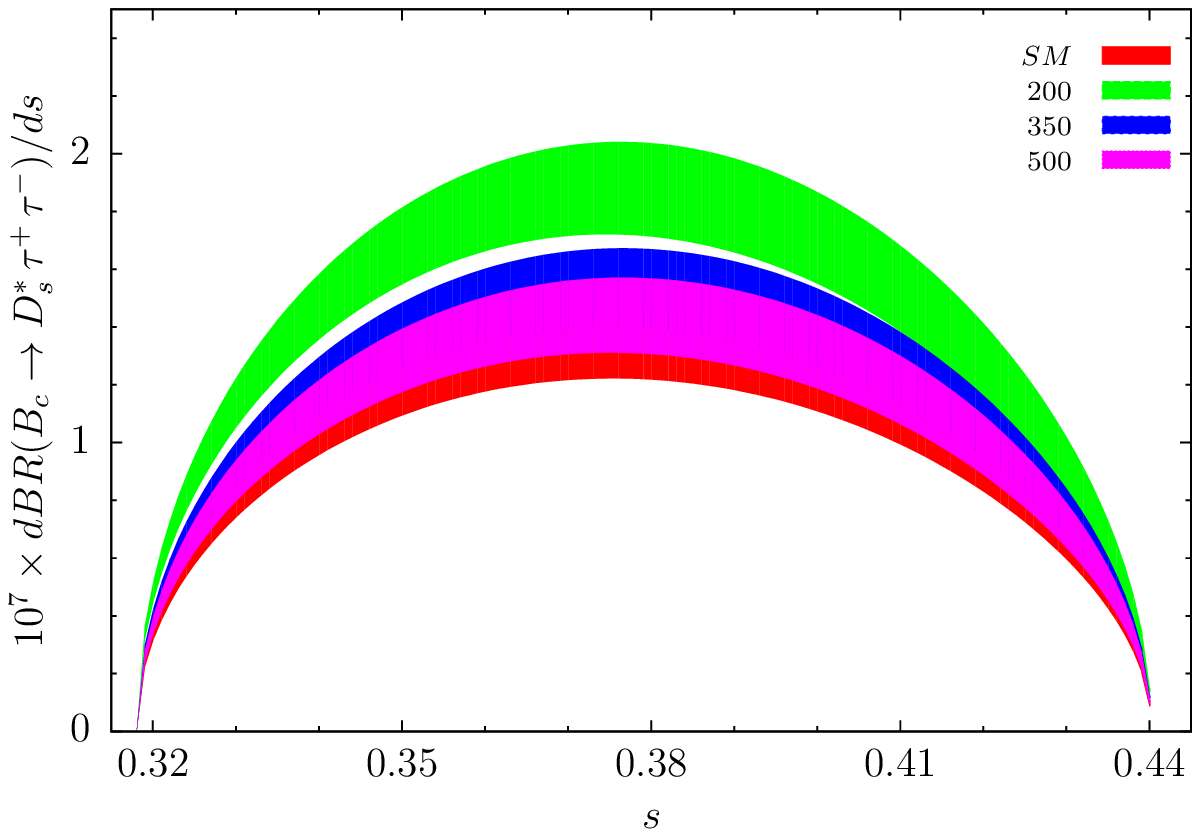}
\caption{ (color online) The dependence of differential branching ratio on s, including the uncertainities on form factors in non-resonance case. (In the legend $1/R=200, 350, 500 \,GeV$.) \label{dbr}}
\end{figure}
\begin{figure}[h]
\centering
\includegraphics[scale=0.62]{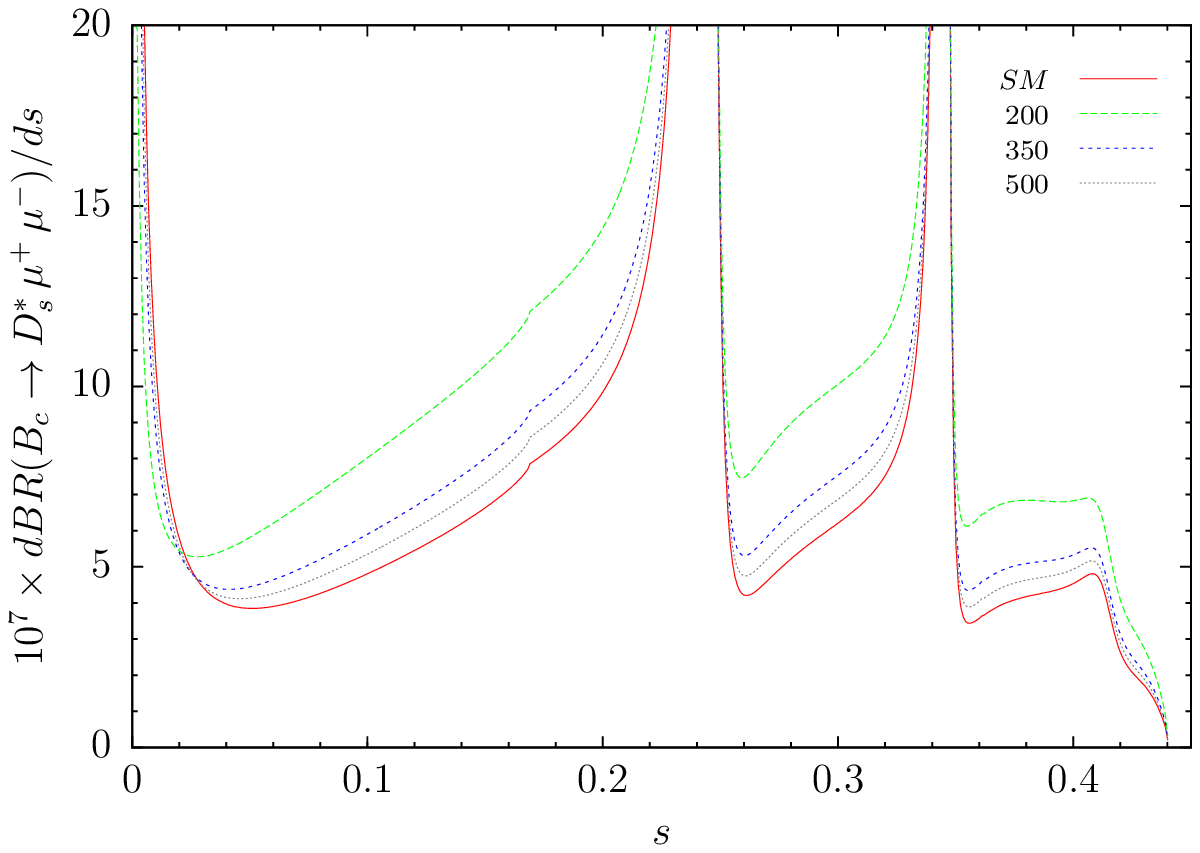}~~~~~
\includegraphics[scale=0.62]{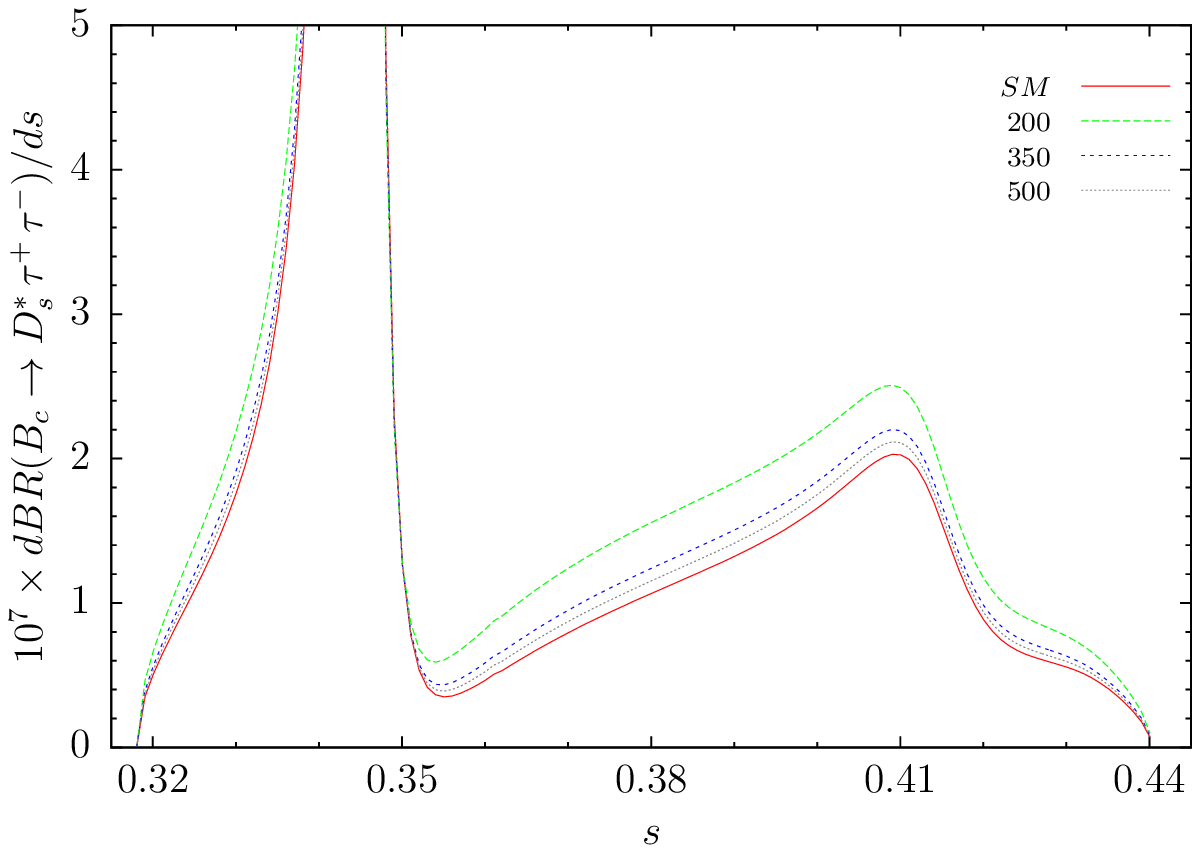}
\caption{ (color online) The dependence of differential branching ratio on s with the central values of form factors including resonance contributions. \label{dbr-res}}
\end{figure}
\begin{figure}[h]
\centering
\includegraphics[scale=0.62]{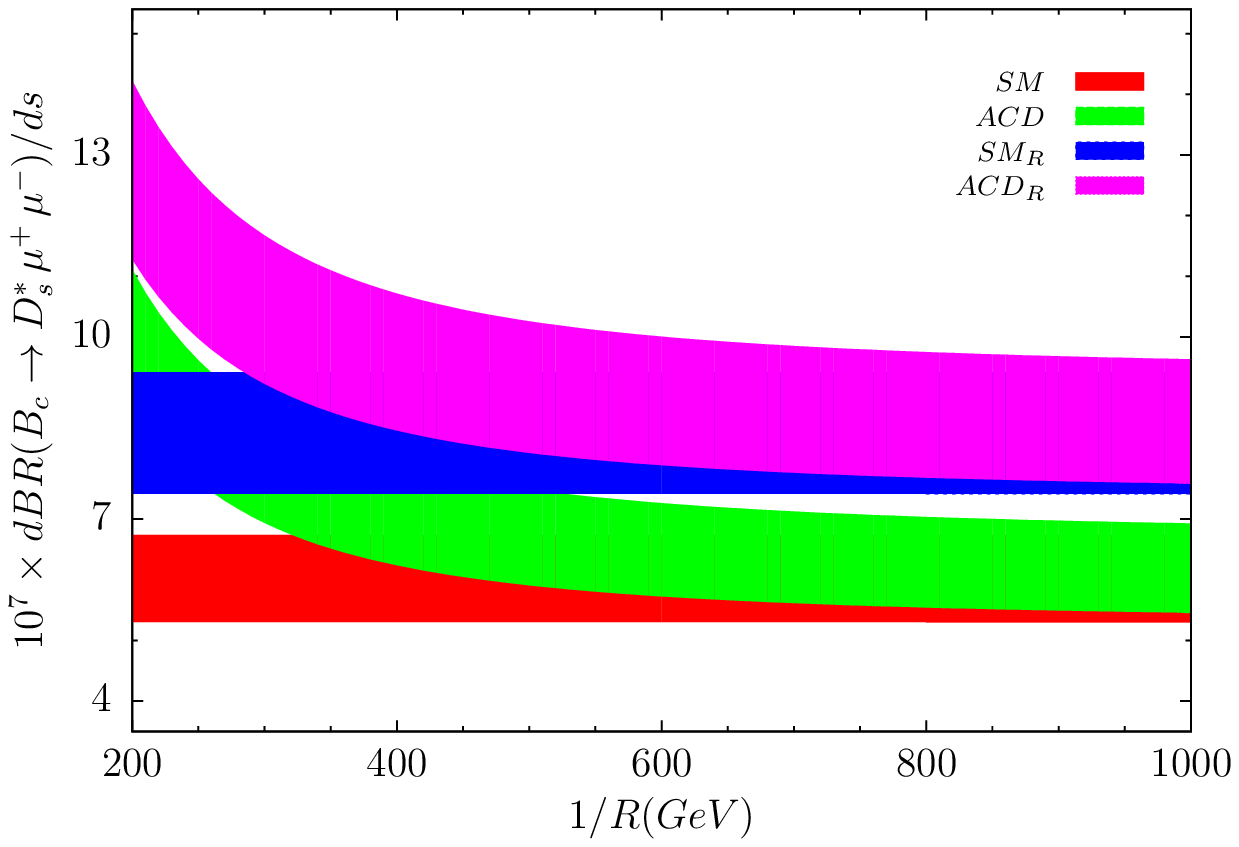}~~~~~
\includegraphics[scale=0.62]{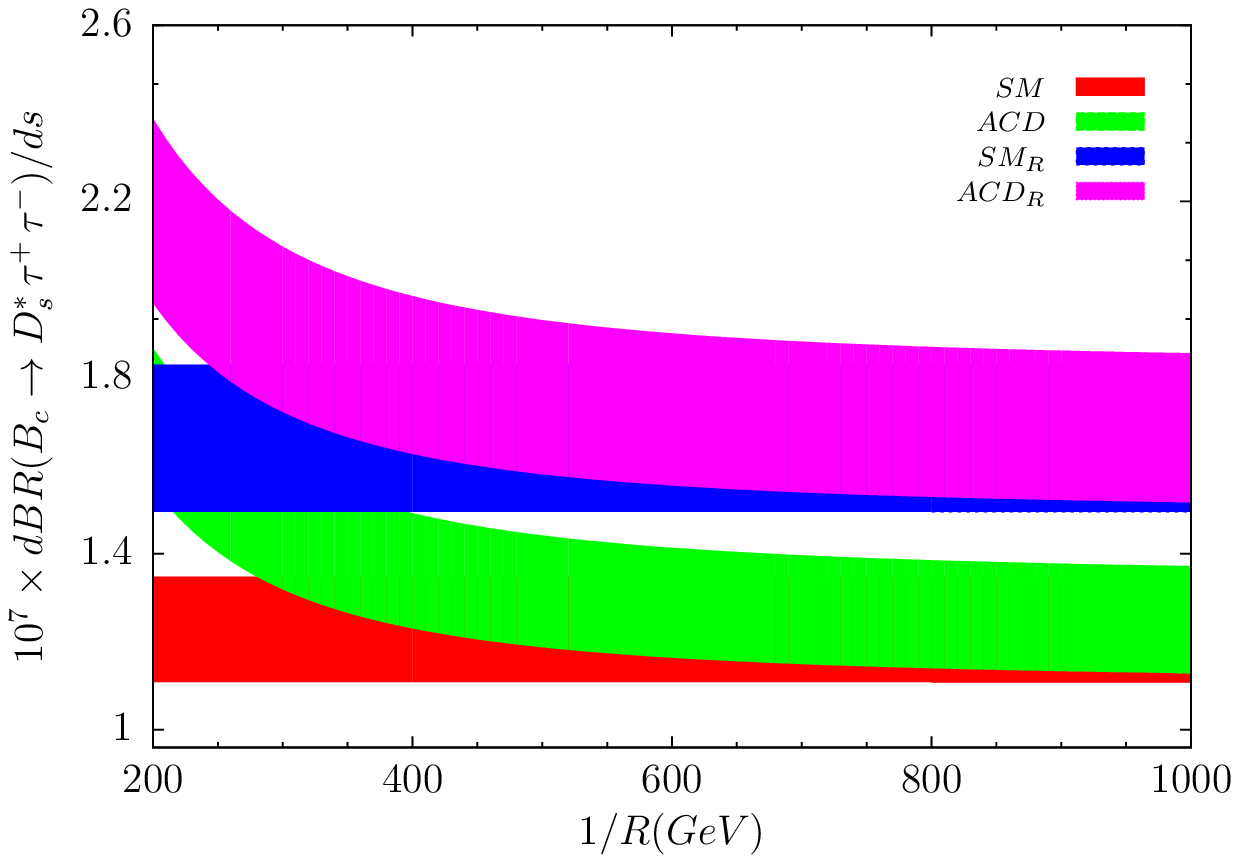}
\caption{ (color online) The dependence of differential branching ratio on 1/R, with and without resonance contributions, including uncertainty on form factors at $s=0.18$ for ${\mu}$, and $s=0.4$ for ${\tau}$. (The subscript R in the legend represents resonance contributions.) \label{dbr-r2}}
\end{figure}

In the numerical analysis, we have used $m_{B_c}=6.28 \, GeV$, $m_{D_{s}^\ast}=2.112 \,GeV$,
$m_b =4.8 \, GeV$, $m_{\mu} =0.105 \, GeV$, $m_{\tau} =1.77 \, GeV$, $|V_{tb} V^*_{ts}|=0.041$,
$G_F=1.17 \times 10^{-5}\, GeV^{-2}$, $\tau_{B_{c}}=0.46 \times 10^{-12} \, s$, and
the values that are not given here are taken from \cite {pdg}. In our work, we have used the numerical values of the form factors calculated in three point QCD sum rules \cite{Azizi08}, in which $q^2$ dependencies of the form factors
are given as
\begin{eqnarray}
F(q^2) = \frac{F(0)}{1+a (q^2/m^{2}_{B_c}) + b (q^2/m^{2}_{B_c})^2}~, \nnb
\end{eqnarray}
and the values of parameters $F(0)$, $a$ and $b$ for the $B_c
\rar D^\ast$ decay are listed in Table \ref{form}.
\begin{table}[h]
\renewcommand{\arraystretch}{1.5}
\addtolength{\arraycolsep}{3pt}
$$
\begin{array}{lccc}
\hline
\hline
& F(0) & a & b \\ \hline
V & \phantom{-} 0.54  \mp 0.018  & -1.28 & -0.230  \\
A_1 & \phantom{-} 0.30  \mp 0.017  & -0.13 & -0.180  \\
A_2 & \phantom{-} 0.36  \mp 0.013  & -0.67 & -0.066  \\
\propto(A_3-A_0) &            -0.57 \mp  0.040  & -1.11 & -0.140  \\
T_1 & \phantom{-} 0.31  \mp 0.017  & -1.28 & -0.230  \\
T_2 & \phantom{-} 0.33  \mp 0.016  & -0.10 & -0.097  \\
T_3 & \phantom{-} 0.29  \mp 0.034  & -0.91 & \phantom{-} 0.007   \\ \hline \hline
\end{array}
$$
\caption{$B_c$ meson decay form factors in the three point QCD sum rules.\label{form}}
\renewcommand{\arraystretch}{1}
\addtolength{\arraycolsep}{-3pt}
\end{table}

The differential branching ratio is calculated without resonance contributions, including uncertainty on form factors, and with resonance contributions, and s dependence for $1/R=200, 350, 500 \,GeV$ is presented in Figs. \ref{dbr} and \ref{dbr-res}, respectively. The change in differential decay rate and difference between the SM results and new effects can be noticed in the figures. The maximum effect is around $s=0.25 \pm0.05 \,(0.37\pm0.02)$ for $\mu\,(\tau)$ in Fig. \ref{dbr}. In spite of the hadronic uncertainty, for $1/R=200\,GeV$ and $350 \,GeV$, studying differential decay rate can be a suitable tool for studying the effect of extra dimension.

Supplementary of these, $1/R$ dependence of differential branching ratio at s= 0.18 (0.4) for $\mu \,(\tau)$ is plotted in Fig. \ref{dbr-r2}.
Considering any given bounds on the compactification factor the effect of universal extra dimension can be seen clearly for low values of R, with and without resonance contributions. On the other hand, $1/R \gsim 600\,GeV$ the contribution varies between $\sim 5-8 \%$ more than the SM results.

To obtain the branching ratio, we integrate Eq. (\ref{bdsunp}) in the allowed physical region. While taking the long-distance contributions into account we introduce some cuts around ${J/ \psi}$ and $\psi(2s)$ resonances to minimize the hadronic uncertainties. The integration region for $q^2$ is divided into three parts for $\mu$ as
$4m_{\mu}^2 \leq q^2 \leq (m_{J/{\psi}}-0.02)^2$,
$(m_{J/{\psi}}+0.02)^2 \leq q^2 \leq (m_{\psi(2s)}-0.02)^2$ and
$(m_{\psi(2s)}+0.02)^2 \leq q^2 \leq (m_{B_c}-m_{D_{s}^\ast})^2$
and for $\tau$ we have
$4m_{\tau}^2 \leq q^2 \leq (m_{\psi(2s)}-0.02)^2$
$(m_{\psi(2s)}+0.02)^2 \leq q^2 \leq (m_{B_c}-m_{D_{s}^\ast})^2$, the same as in \cite{Gursevil02}.

The results of branching ratio in the SM with resonance contributions and uncertainty on form factors, we obtain
\bea
&&Br(B_c \rar D_s^{*} \mu^+ \mu^-)=2.13^{+0.27}_{-0.25}\times 10^{-7}\nnb \\
&&Br(B_c \rar D_s^{*} \tau^+ \tau^-)=1.45^{+0.15}_{-0.14}\times 10^{-8}.
\eea
\begin{figure}[h]
\centering
\includegraphics[scale=0.62]{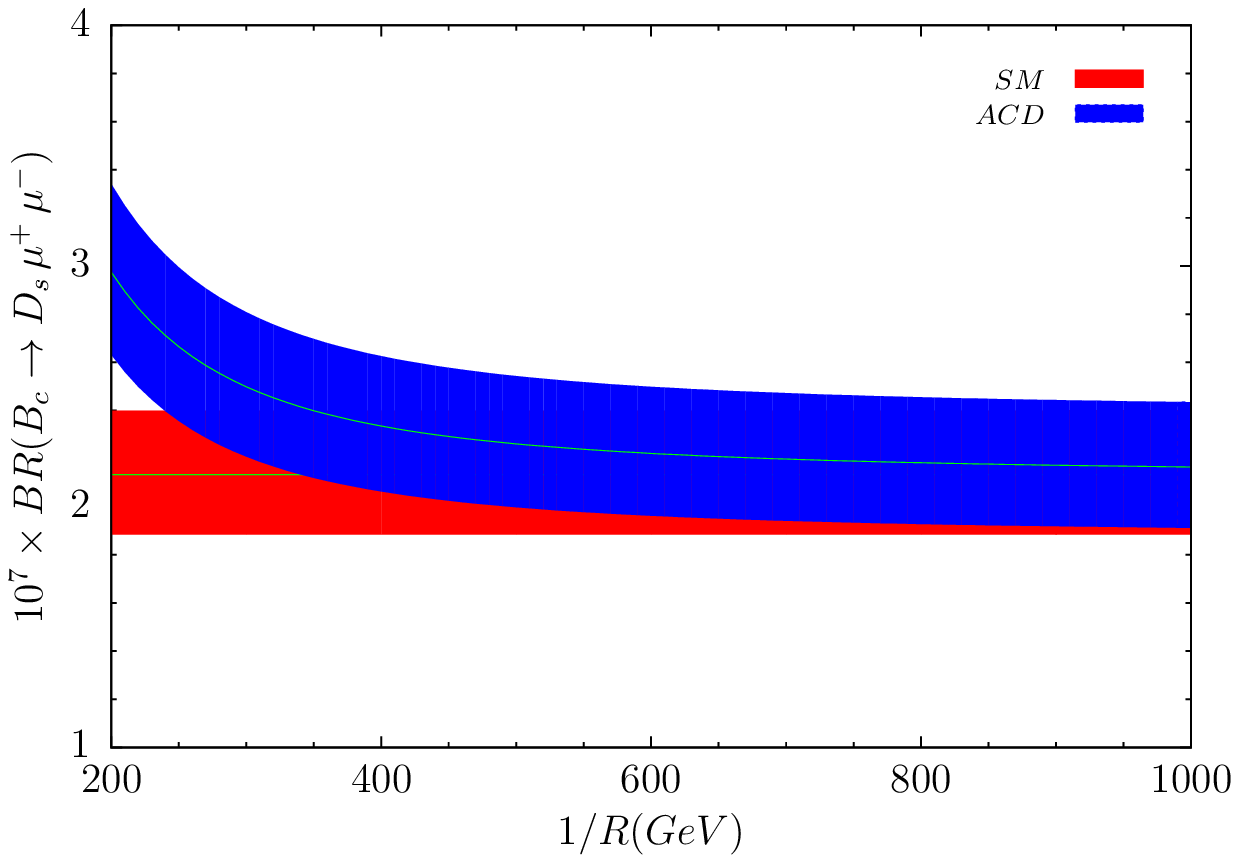}~~~~~
\includegraphics[scale=0.62]{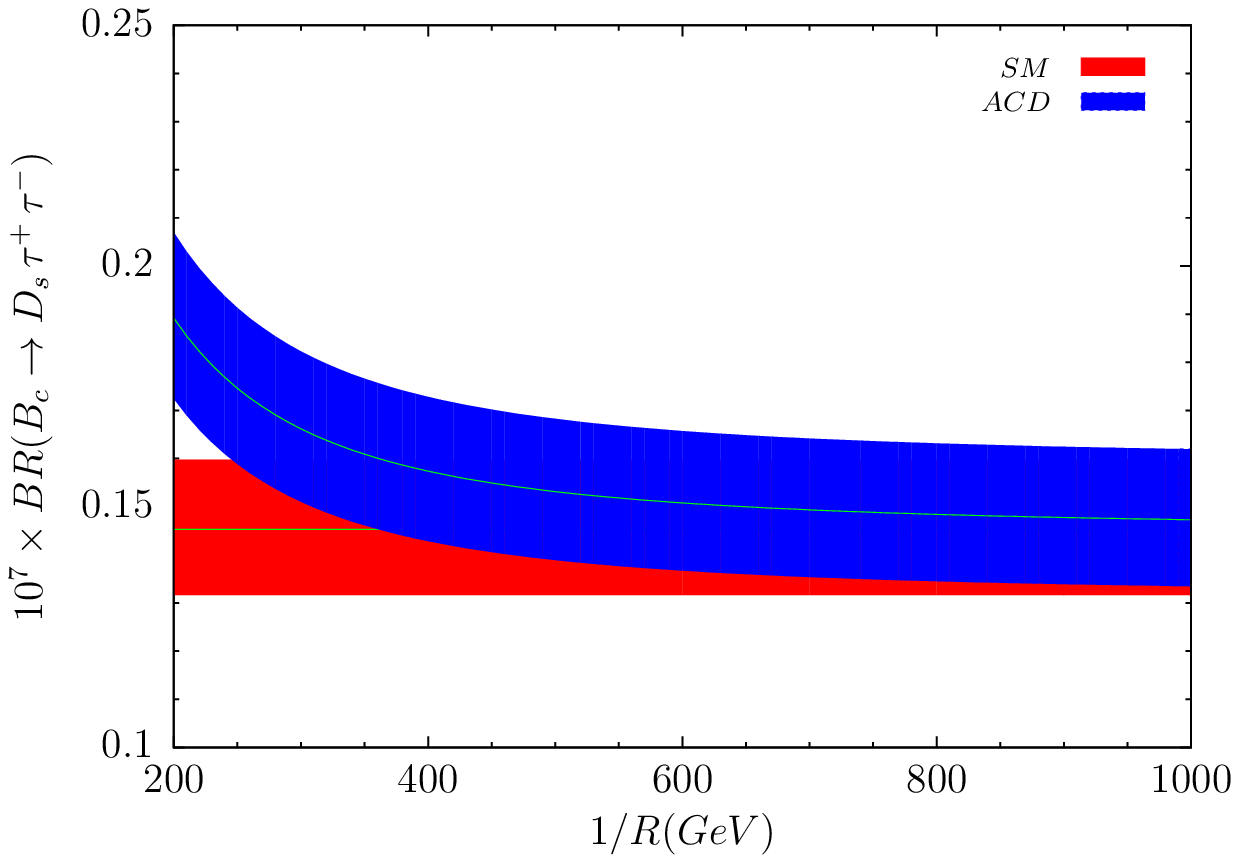}
\caption{(color online) The dependence of branching ratio on $1/R$, including resonance contributions and uncertainty on form factors. \label{br}}
\end{figure}
Observing the contribution of the ACD, the $1/R$ dependent branching ratios, including resonance contributions and uncertainty on form factors, are given in Fig. \ref{br}.
Comparing the SM results and our theoretical predictions on the branching ratio for both decay channels, the lower bound for $1/R$ is found approximately $250\,GeV$ which is consistent with the previously mentioned results.

As $1/R$ increases, the branching ratios approach to their SM values. For $1/R \geq  550\, GeV$ in both channels, they become less than $5\%$ greater than that of the SM values.
Between $1/R=250-350\, GeV$ the ratio is $(2.66-2.40)^{+0.30}_{-0.28}\times 10^{-7}$ for ${\mu}$, $(1.75-1.61)^{+0.16}_{-0.15}\times 10^{-8}$ for $\tau$ decay. Comparing these with the SM results, the differences worth to study and can be considered as a signal of new physics and an evidence of existence of extra dimension.
\section{Forward-Backward Asymmetry}
Another efficient tool for establishing new physics is the study of forward-backward asymmetry. The position of zero value of $A_{FB}$ is very sensitive to the new physics.
The normalized differential form is defined for final state leptons as
 \bea
\label{fbasimetri}  A_{FB}(s) &=&
            \frac{\int_{0}^{1}\frac{d^2\Gamma}{ds
            dz} dz-\int_{-1}^{0}\frac{d^2\Gamma}{ds
            dz}dz}
            {\int_{0}^{1}\frac{d^2\Gamma}{ds
            dz}dz+\int_{-1}^{0}\frac{d^2\Gamma}{ds
            dz}dz}\nnb \\
\eea
where $z=\cos\theta$ and $\theta$ is the angle between the directions of $\ell^{-}$ and
$B_c$ in the rest frame of the lepton pair.

\begin{figure}[h]
\centering
\includegraphics[scale=0.62]{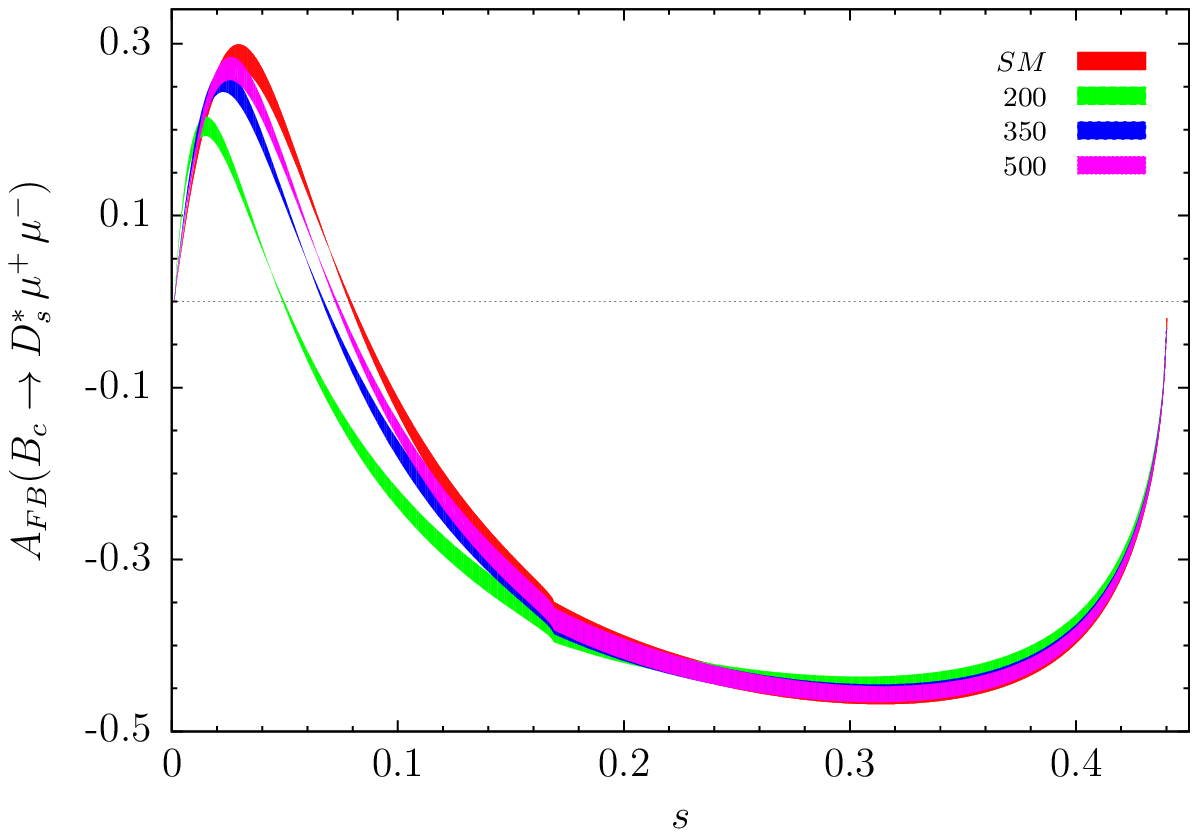}~~~~~
\includegraphics[scale=0.62]{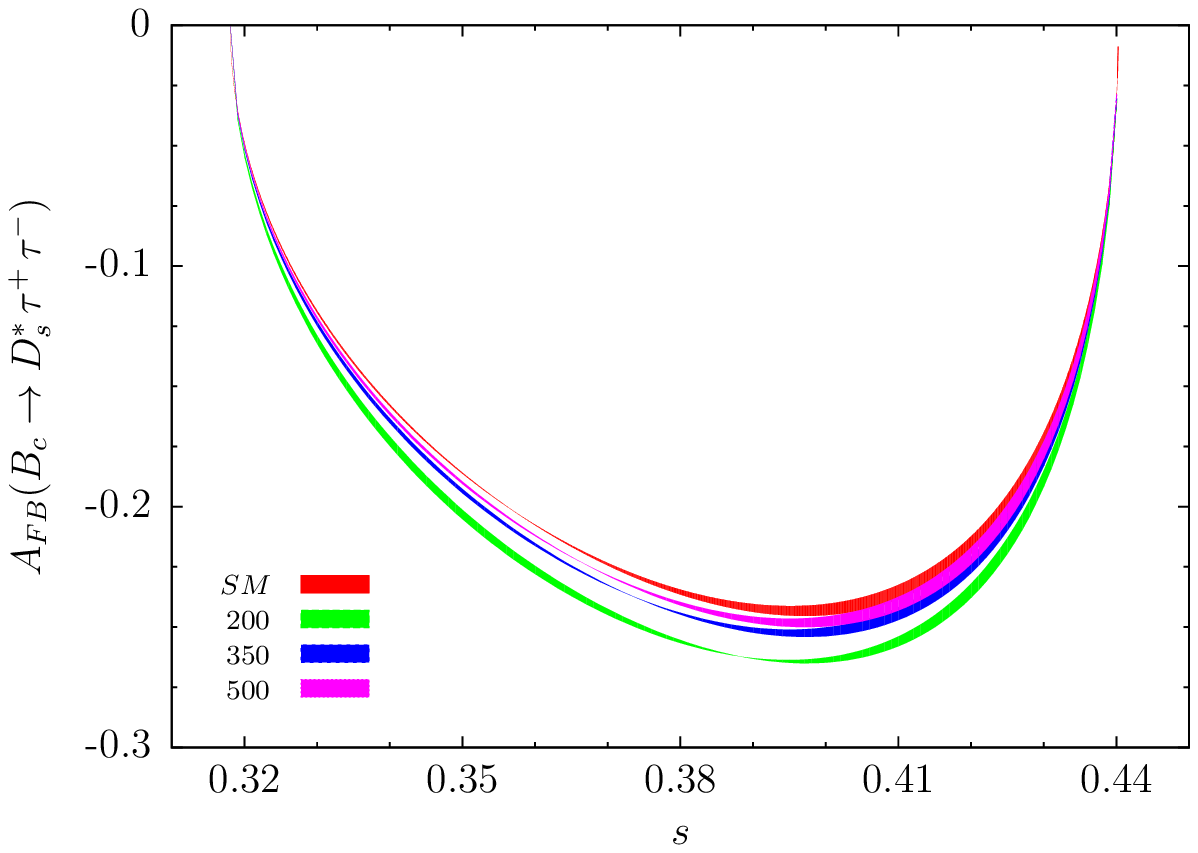}
\caption{(color online) The lepton forward-backward asymmetry including uncertainty on form factors. \label{afb}}
\end{figure}

\begin{figure}[h]
\centering
\includegraphics[scale=0.62]{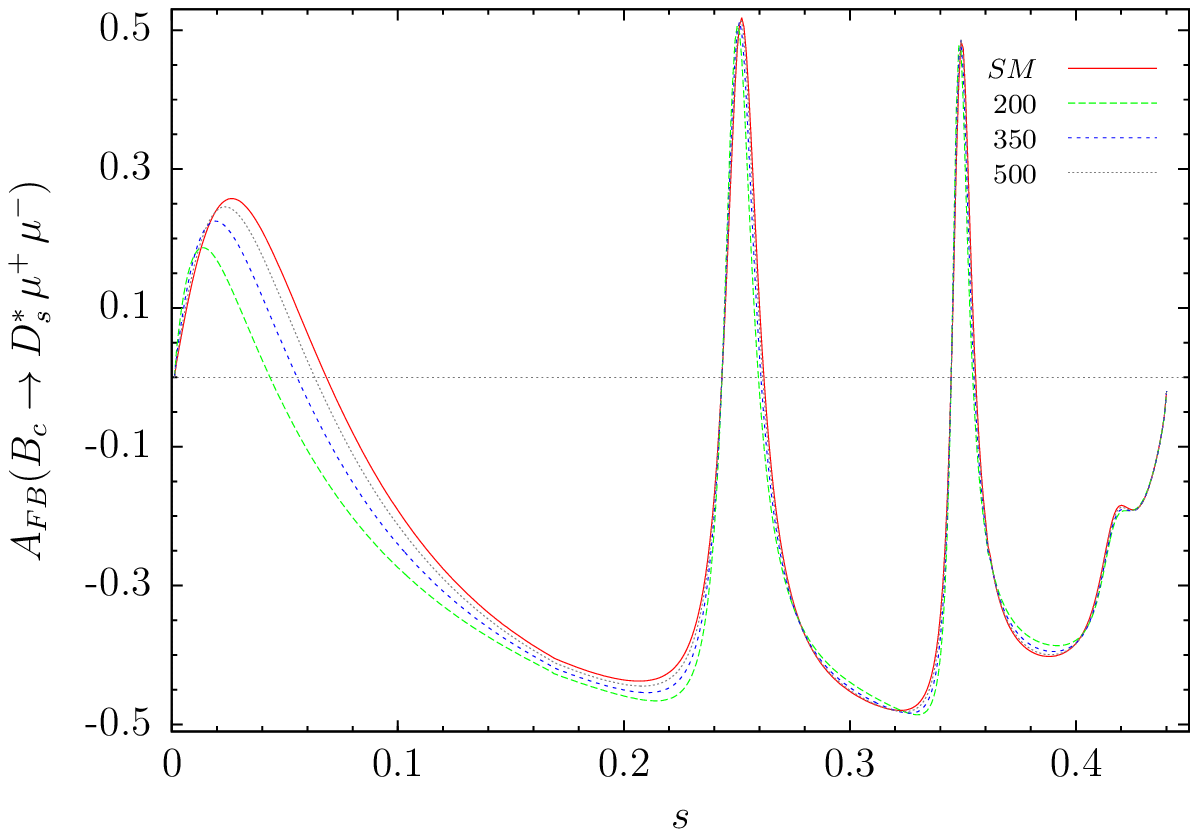}~~~~~
\includegraphics[scale=0.62]{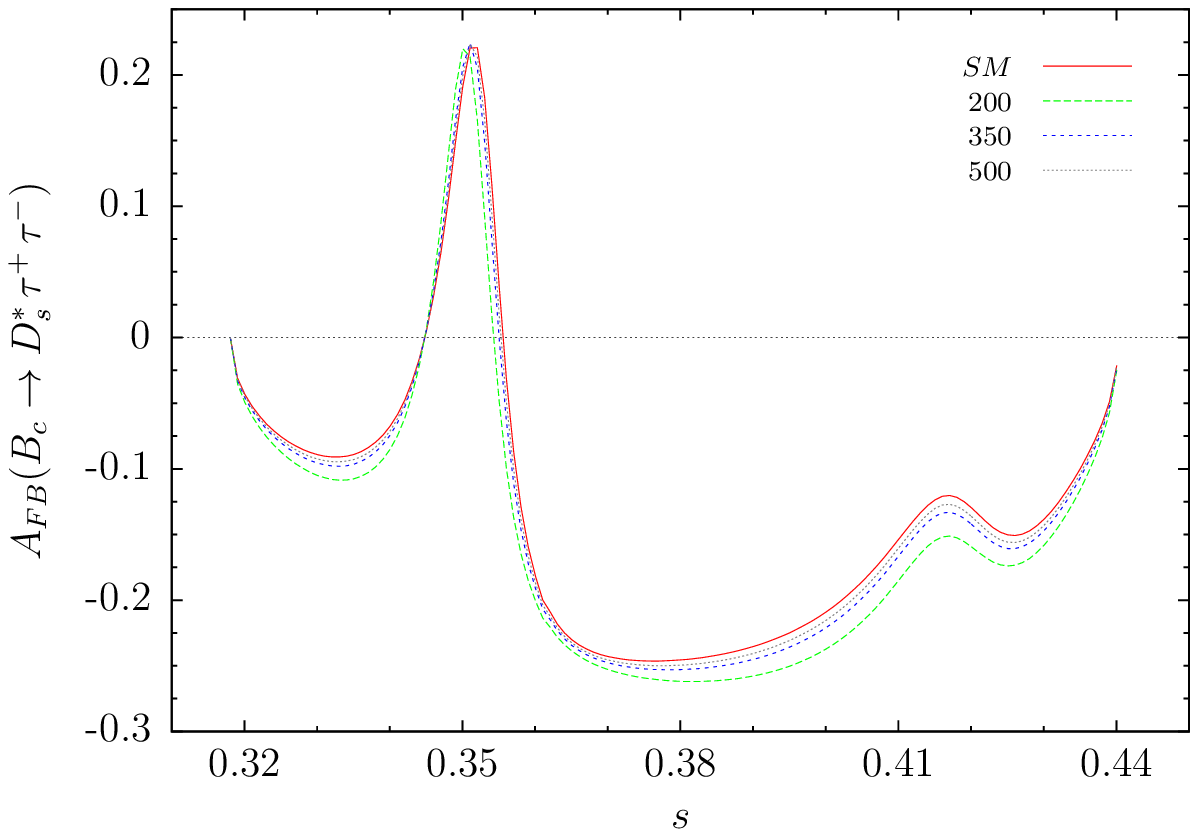}
\caption{(color online) The lepton forward-backward asymmetry including resonance contributions \label{afb-res}}
\end{figure}

In the case of \BcDsll, we get
\bea
\label{fba} A_{FB}
&=&    \frac{G^2 \alpha^2
        m_{B_c}}{2^{12} \pi^5 }
         \vel V_{tb} V_{ts}^\ast \ver^2
         \frac{8m_{B_c}^4\sqrt{\lambda}v s
         (Re[BE^\ast]+Re[AF^\ast])}{d\Gamma/ds}  \nnb \\
&=& \frac {8m_{B_c}^4 \sqrt{\lambda}v s (Re[BE^\ast]+Re[AF^\ast])}
   {\Delta_{D_{s}^\ast }}.
 \eea
Using above equation, we present the variation of lepton forward-backward asymmetry with s including uncertainty on form factors in Fig. \ref{afb}. As $1/R$ gets smaller, there appears considerable difference between the SM and the ACD results for $s\lsim0.16$ in $\mu$  and $0.33\lsim s\lsim 0.43$ in $\tau$ decays.
Considering the resonance contributions, the results are given in Fig. \ref{afb-res}, one can recognize a similar situation for $s\lsim 0.23$ and $0.32\lsim s\lsim 0.44$, respectively.

\begin{figure}[h]
\centering
\includegraphics[scale=0.62]{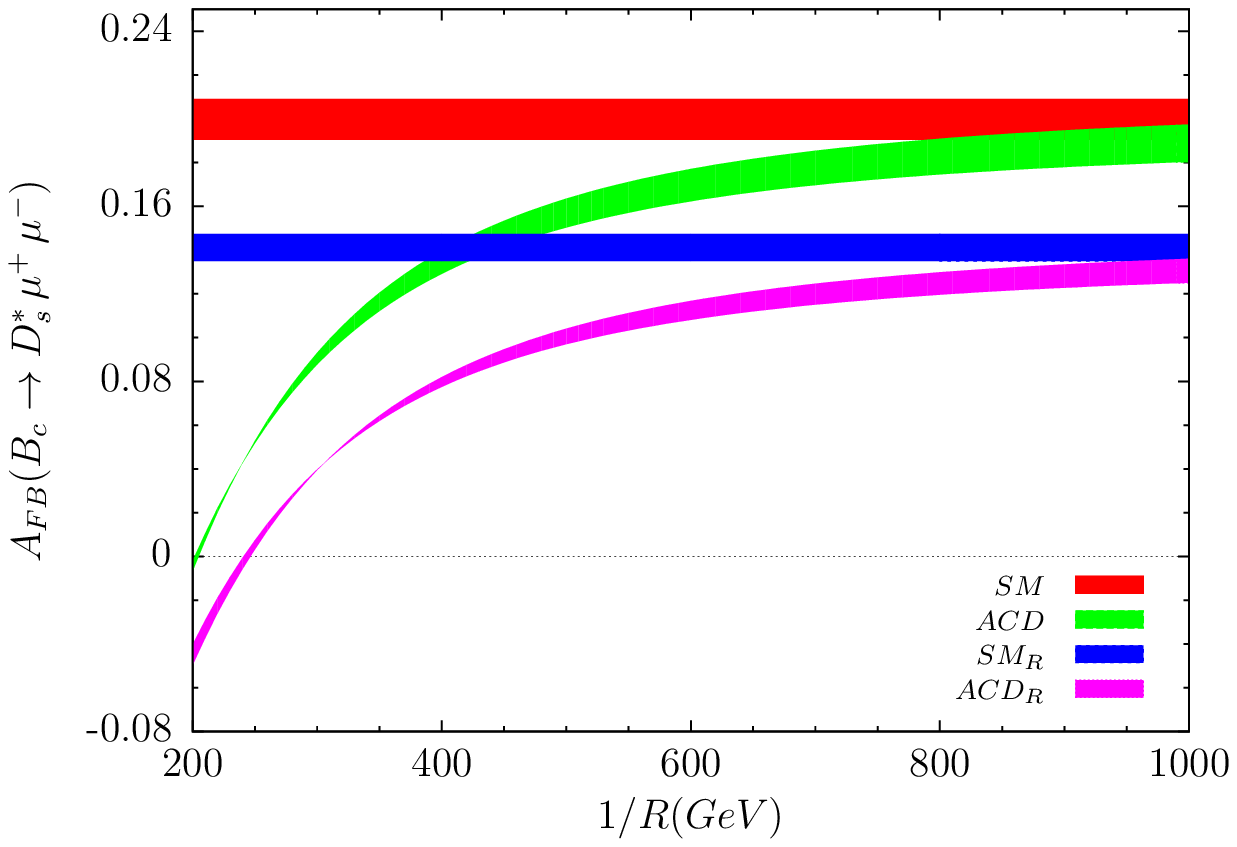}~~~~~
\includegraphics[scale=0.62]{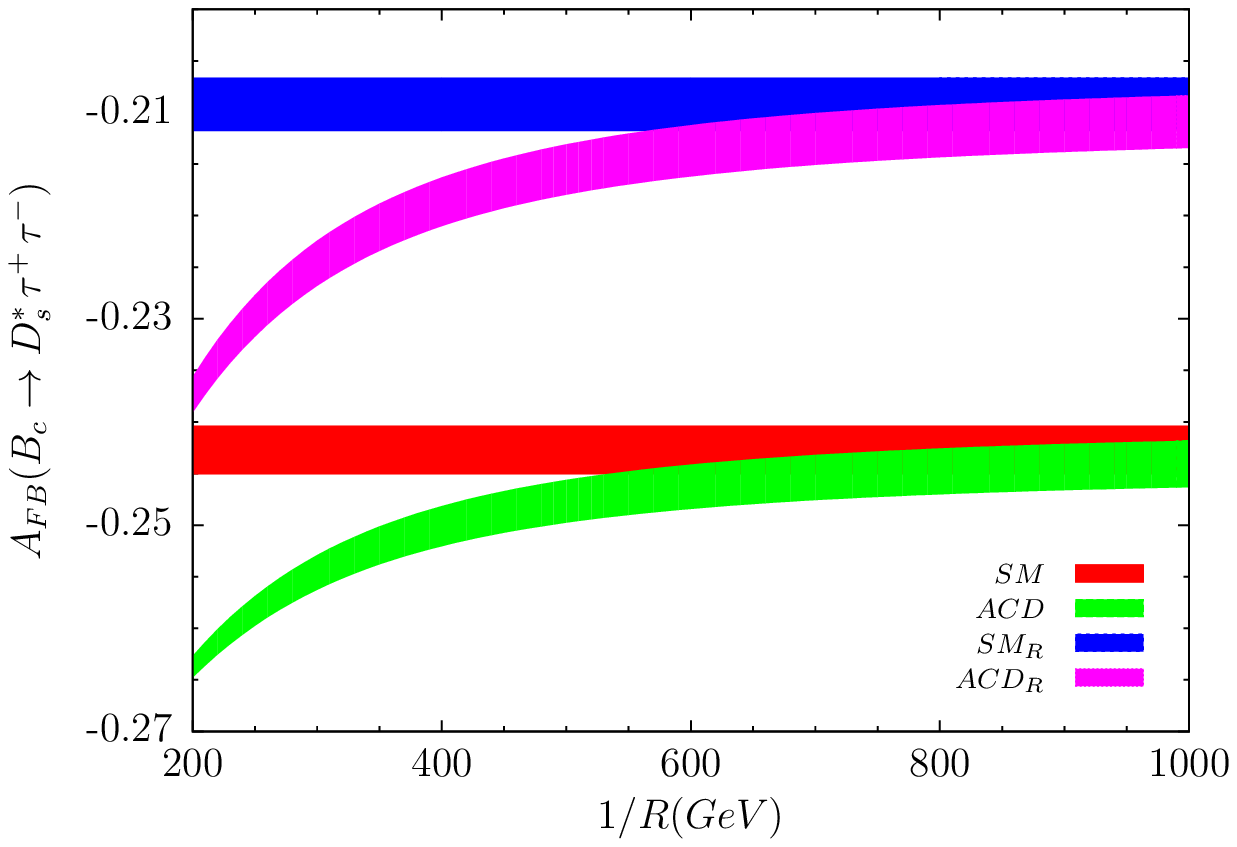}
\caption{(color online) The dependence of lepton forward-backward asymmetry
on $1/R$ at $s=0.05$ for ${\mu}$ and $s=0.4$ for ${\tau}$.  \label{afb-R}}
\end{figure}

To understand the dependence of $A_{FB}$ on $1/R$ for both lepton channels better, we perform calculation at $s=0.05 \,(0.4)$ for $\mu (\tau)$ and present the results in Fig. \ref{afb-R}. In the $\mu$ channel, UED contribution on $A_{FB}$ gets important between $1/R=200-600\,GeV$, while in $\tau$ decay the contribution is insignificant for $1/R\gsim 400\,GeV$.

\begin{figure}[h]
\centering
\includegraphics[scale=0.62]{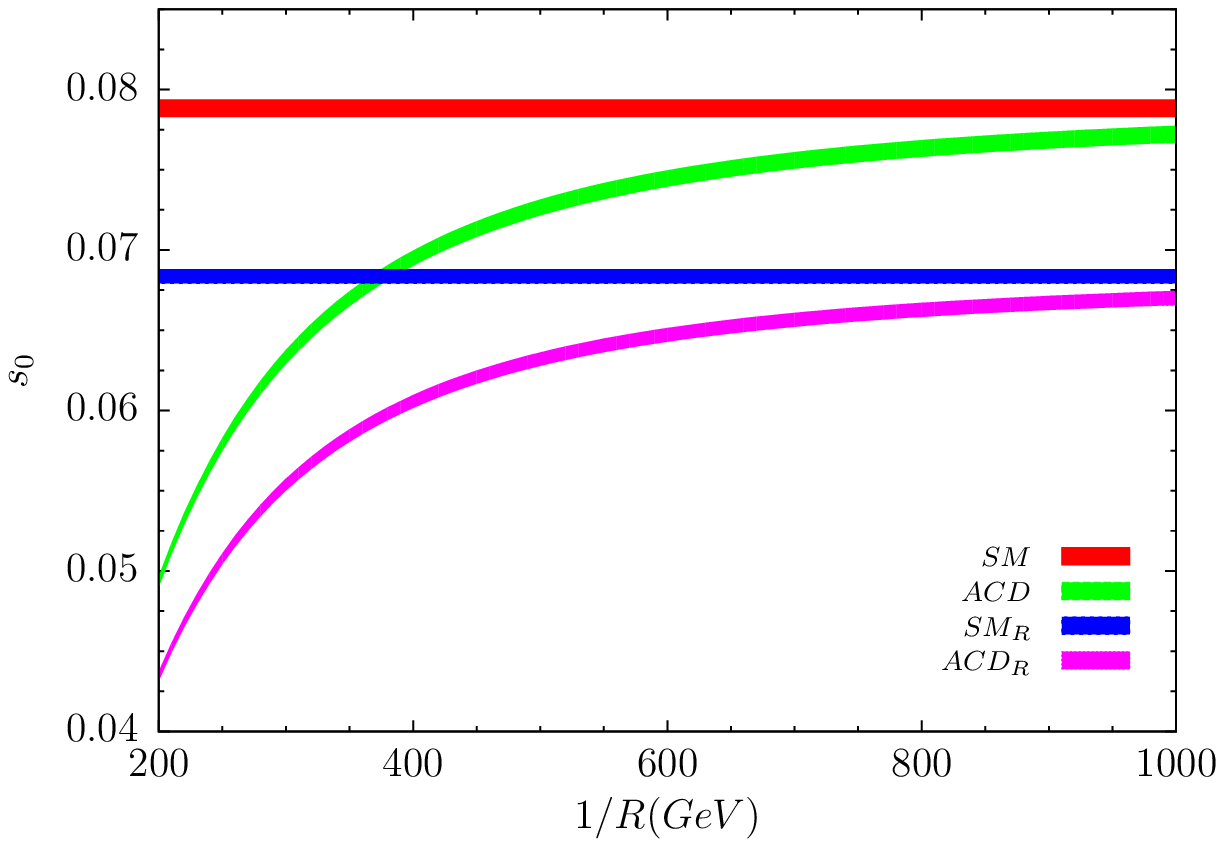}~~~~~
\includegraphics[scale=0.62]{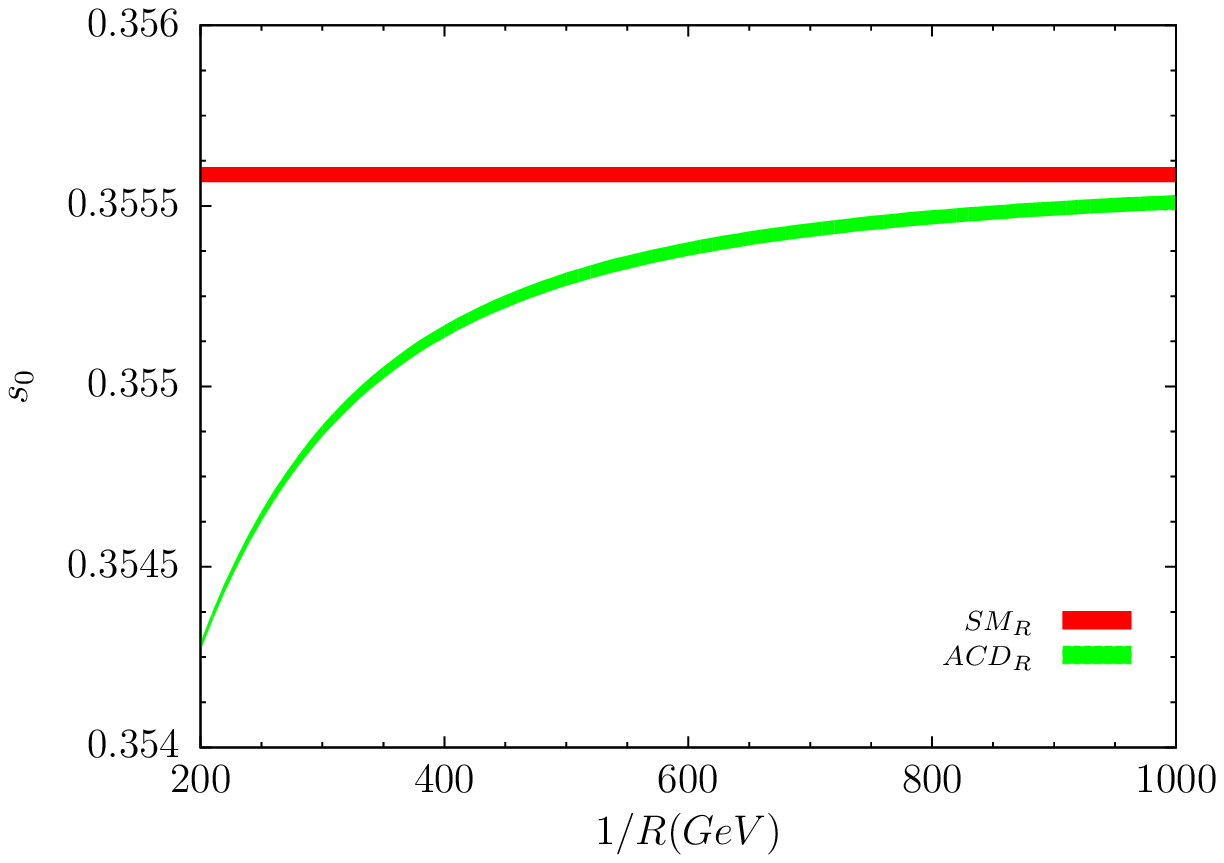}
\caption{(color online) The variation of zero position of lepton forward-backward asymmetry with $1/R$. \label{afb-zero}}
\end{figure}

The position of the zero of forward-backward asymmetry, $s_0$, is determined numerically and the results are presented in Fig. \ref{afb-zero}. Both plots for \tepm is for the zero point in the $s<0.1$ region; the lower (upper) one is for the resonance (non resonance) case, while the zero point for \tept is because of resonance contributions.
In the SM, resonance shifts the zero point of the asymmetry, $s_0=0.079$, to a lower value, $s_0=0.068$, in \tepm, i.e., further corrections could shift $s_0$ to smaller values \cite{Colangelo06}.
As $1/R\rightarrow 200\, GeV$ the $s_0$ approaches low values for both decay channels.
In the $1/R=250-350\,GeV$ region, $s_0$ varies between ($0.058-0.068$) without resonance contributions and ($0.051-0.058$) with resonance contributions. The $s_0$ shift is $\sim 5\%$ of the SM value for $1/R\gsim600 \,GeV$. The variation of $s_0$ for \tept is negligible.

\section{Lepton Polarization Asymmetries}

We will discuss the possible effects of the ACD model in lepton polarization, a way of searching new physics.
Using the convention followed by previous works \cite{Fukae2000}-\cite{Kruger96}, in the rest frame of $\ell^-$ we define the orthogonal unit vectors $S_i^-$, for the polarization of the lepton along the longitudinal,
transverse and normal directions as
\bea
\label{pol}
S_L^{-} &\equiv& (0,\vec{e}_L) =
    \ga 0,\frac{\vec{p}_{\ell}}{\vel \vec{p}_{\ell} \ver} \dr, \nnb \\
S_N^{-} &\equiv& (0,\vec{e}_N) =
    \ga 0,\frac{\vec{p}_{D_{s}^\ast} \times \vec{p}_{\ell}}
    {\vel \vec{p}_{D_{s}^\ast} \times \vec{p}_{\ell} \ver} \dr, \nnb \\
S_T^{-} &\equiv& (0,\vec{e}_T) =
    \ga 0, \vec{e}_N \times \vec{e}_L \dr,
\eea
where $\vec{p}_{\ell}$ and $\vec{p}_{D_{s}^\ast}$ are the three momenta of $\ell^-$ and
$D_{s}^\ast$ meson in the center of mass (CM) frame of $\ell^+ \ell^-$
system, respectively. The longitudinal unit vector $S_L^-$ is
boosted by Lorentz transformation,
\bea
\label{bs}
S^{-\mu}_{L,\, CM} = \ga \frac{\vel \vec{p}_{\ell} \ver}{m_\ell},
    \frac{E_\ell \,\vec{p}_{\ell}}{m_\ell \vel \vec{p}_{\ell} \ver} \dr,
\eea
while vectors of perpendicular directions remain unchanged under the Lorentz boost.

The differential decay rate of $B_c \rar D_s^{\ast} \ell^+ \ell^-$ for
any spin direction $\vec{n}^{-}$ of the $\ell^{-}$
can be written in the following form
\bea
\label{ddr}
\frac{d\Gamma(\vec{n}^{-})}{ds} = \frac{1}{2}
\ga \frac{d\Gamma}{ds}\dr_0
\Bigg[ 1 + \Bigg( P_L^{} \vec{e}_L^{\,-} + P_N^{-}
\vec{e}_N^{\,-} + P_T^{-} \vec{e}_T^{\,-} \Bigg) \cdot
\vec{n}^{-} \Bigg]\,.
\eea
Here, $(d \Gamma / ds)_0$ corresponds to the unpolarized decay
rate, whose explicit form is given in Eqn. (\ref{bdsunp}).

The polarizations $P^{-}_L$, $P^{-}_T$ and $P^{-}_N$ in Eq. (\ref{ddr}) are defined by the equation
\bea P_i^{-}(s) = \frac{\ds{\frac{d \Gamma}{ds}
                   ({\bf{n}}^{-}={\bf{e}}_i^{\,-}) -
                   \frac {d \Gamma}{ds}
                   ({\bf{n}}^{-}=-{\bf{e}}_i^{\,-})}}
             {\ds{\frac{d \Gamma}{ds}
             ({\bf{n}}^{-}={\bf{e}}_i^{\,-}) +
             \frac{d \Gamma}{ds}
             ({\bf{n}}^{-}=-{\bf{e}}_i^{\,-})}}~, \nnb
\eea
for $i=L,~N,~T$. Here, $P^{-}_L$ and $P^{-}_T$ represent the longitudinal and transversal asymmetries, respectively, of the charged lepton $\ell^{-}$ in the
decay plane, and $P^{-}_N$ is the normal component to both of them.

The explicit form of longitudinal polarization for $\ell^-$ is
\bea
\label{bdlong}
P^-_L&=&\frac{1}{3\Delta_{D^\ast_s}} 4 m^2_{B_c} v \Big[
    8 m^4_{B_c} s \lambda Re[AE^\ast]+\frac{1}{r}(12rs+\lambda)Re[BF^\ast] \nnb \\
& & -\frac{1}{r}\lambda m^2_{B_c}(1-r-s)\Big[Re[BG^\ast]+ Re[CF^\ast] \Big]
        + \frac {1}{r} {\lambda^2} m^4_{B_c} Re[CG^\ast]
\Big].
\eea
\begin{figure}[h]
\centering
\includegraphics[scale=0.62]{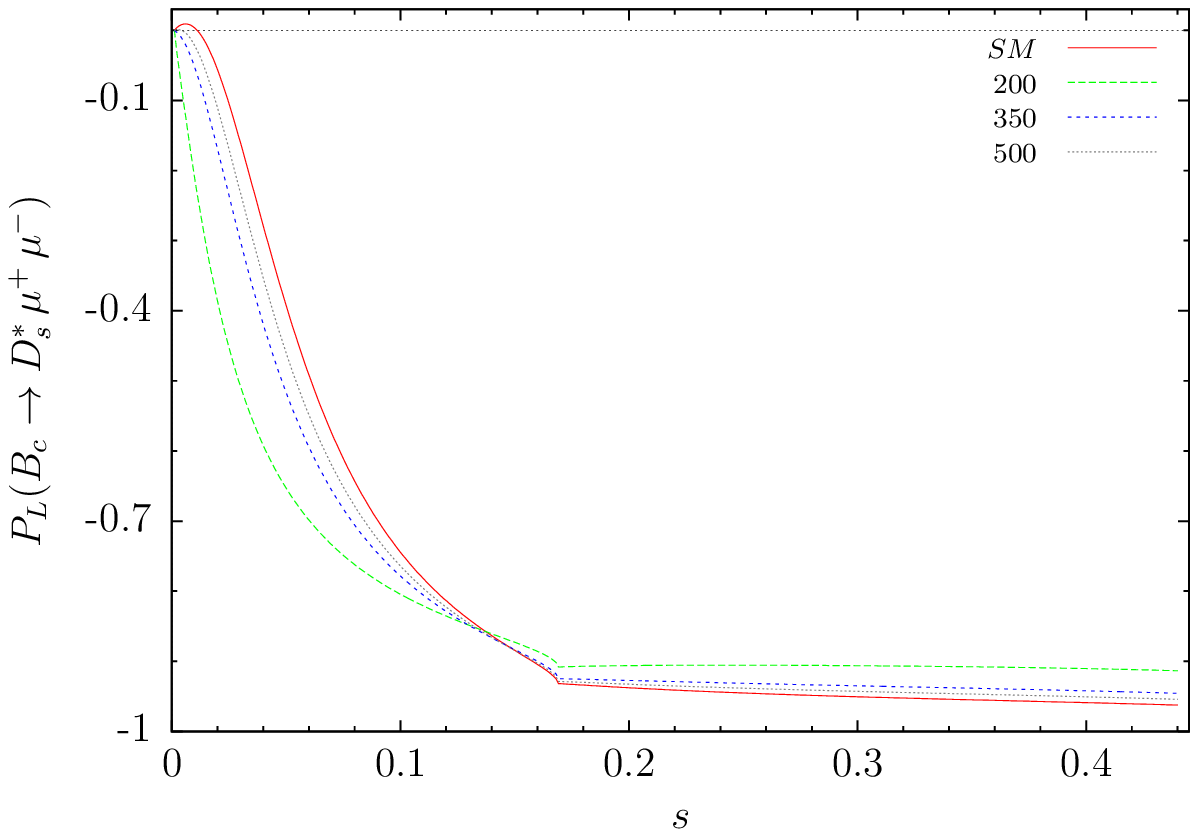}~~~~~
\includegraphics[scale=0.62]{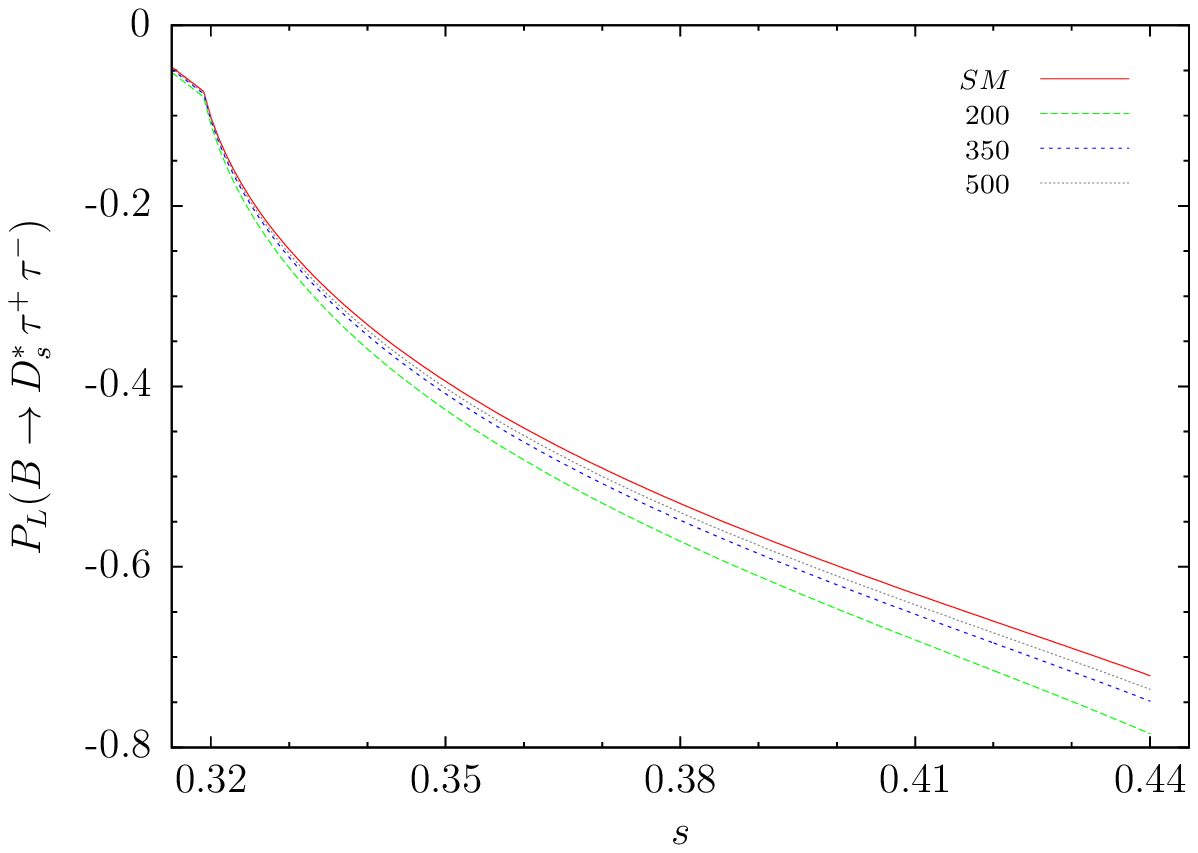}
\caption{(color online) The dependence of longitudinal polarization on s without resonance contributions using central values of form factors. \label{LongPol-ss}}
\end{figure}
\begin{figure}[h]
\centering
\includegraphics[scale=0.62]{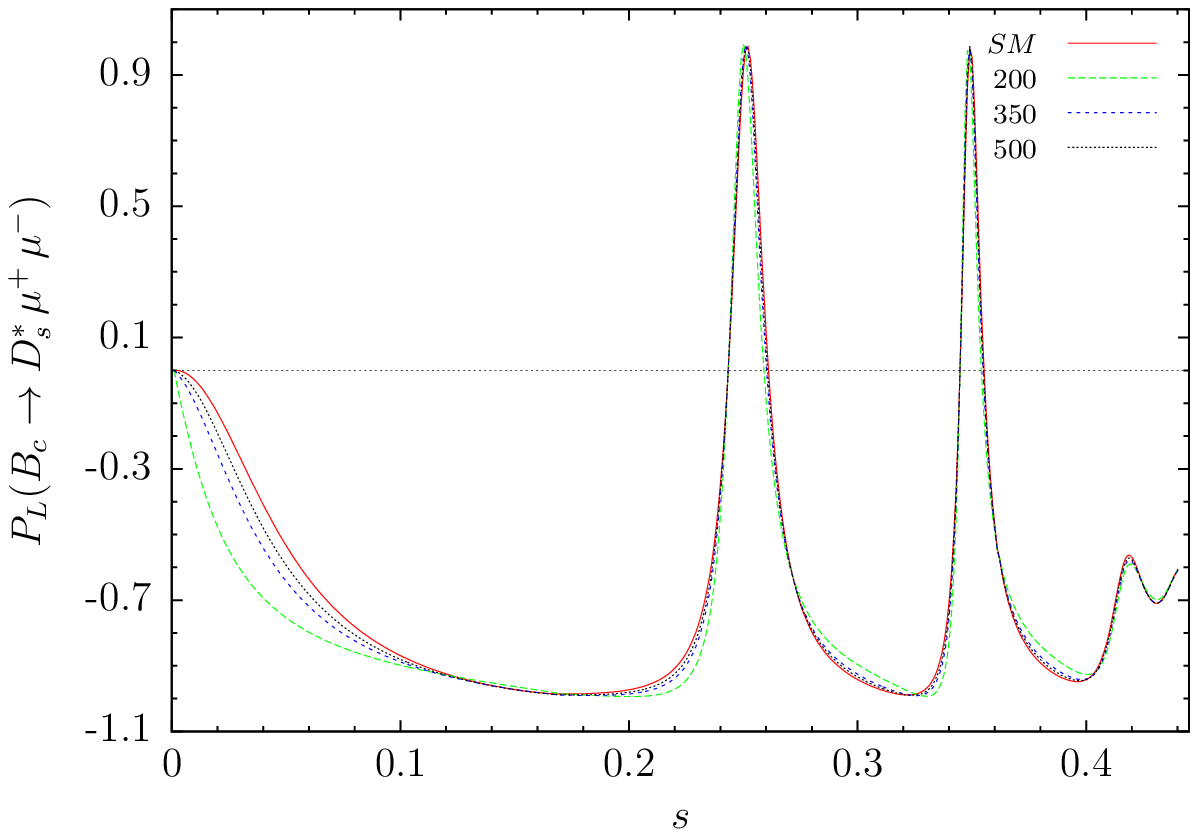}~~~~~
\includegraphics[scale=0.62]{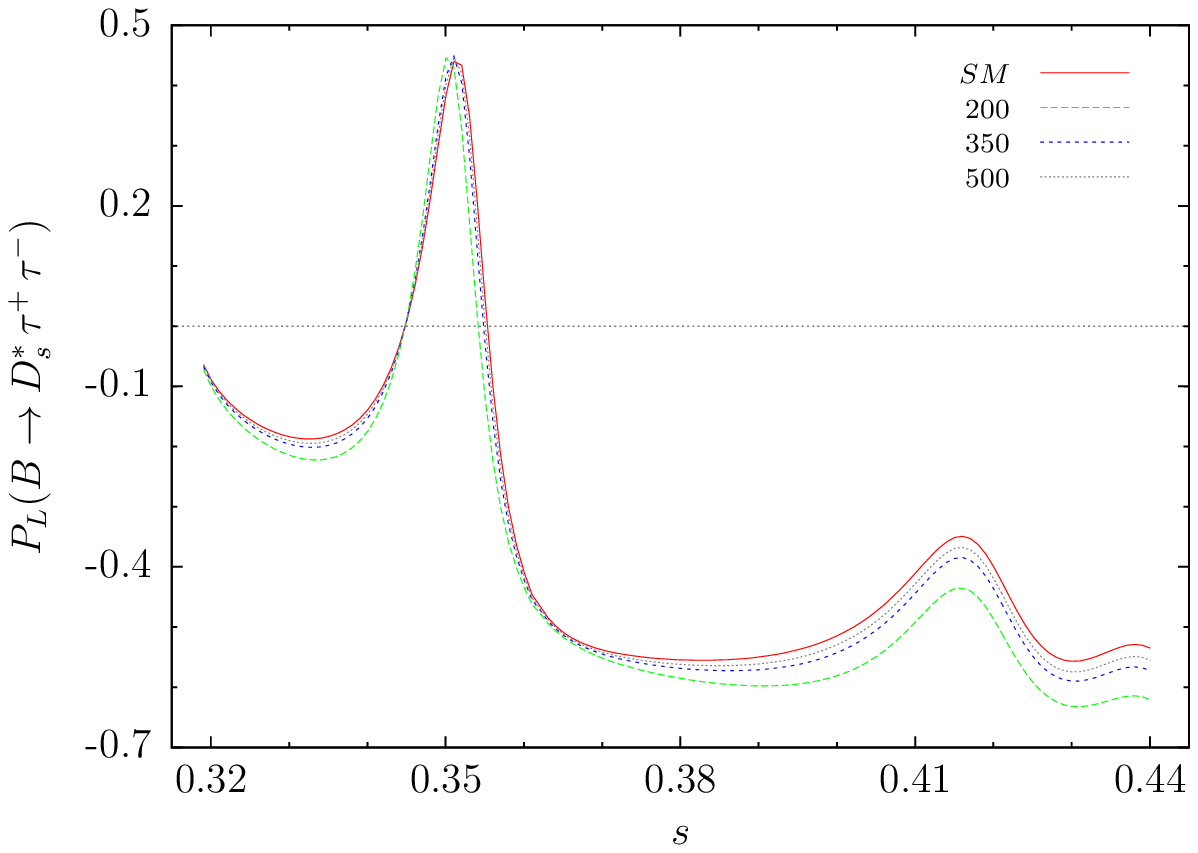}
\caption{(color online) The dependence of longitudinal polarization on s with resonance contributions using central values of form factors. \label{LongPol-res-ss}}
\end{figure}

\begin{figure}[h]
\centering
\includegraphics[scale=0.62]{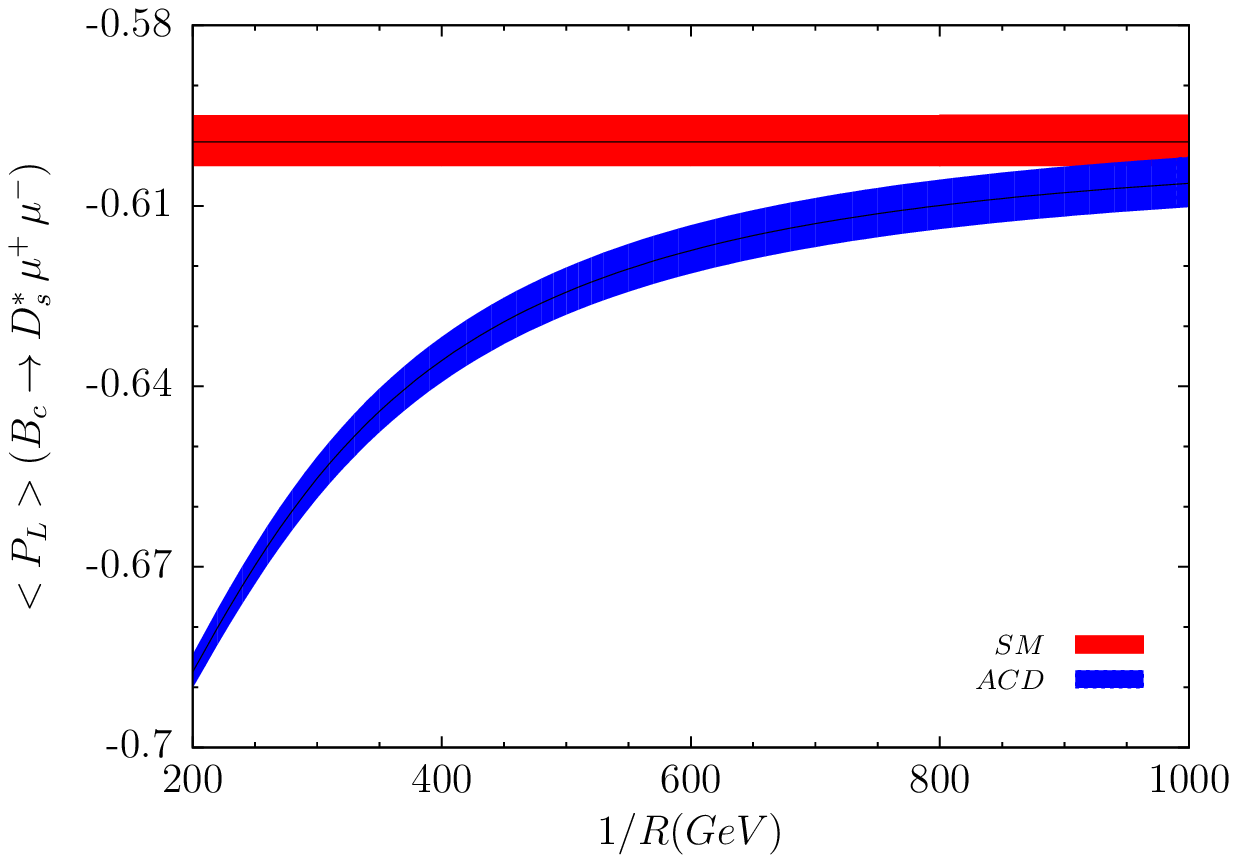}~~~~~
\includegraphics[scale=0.62]{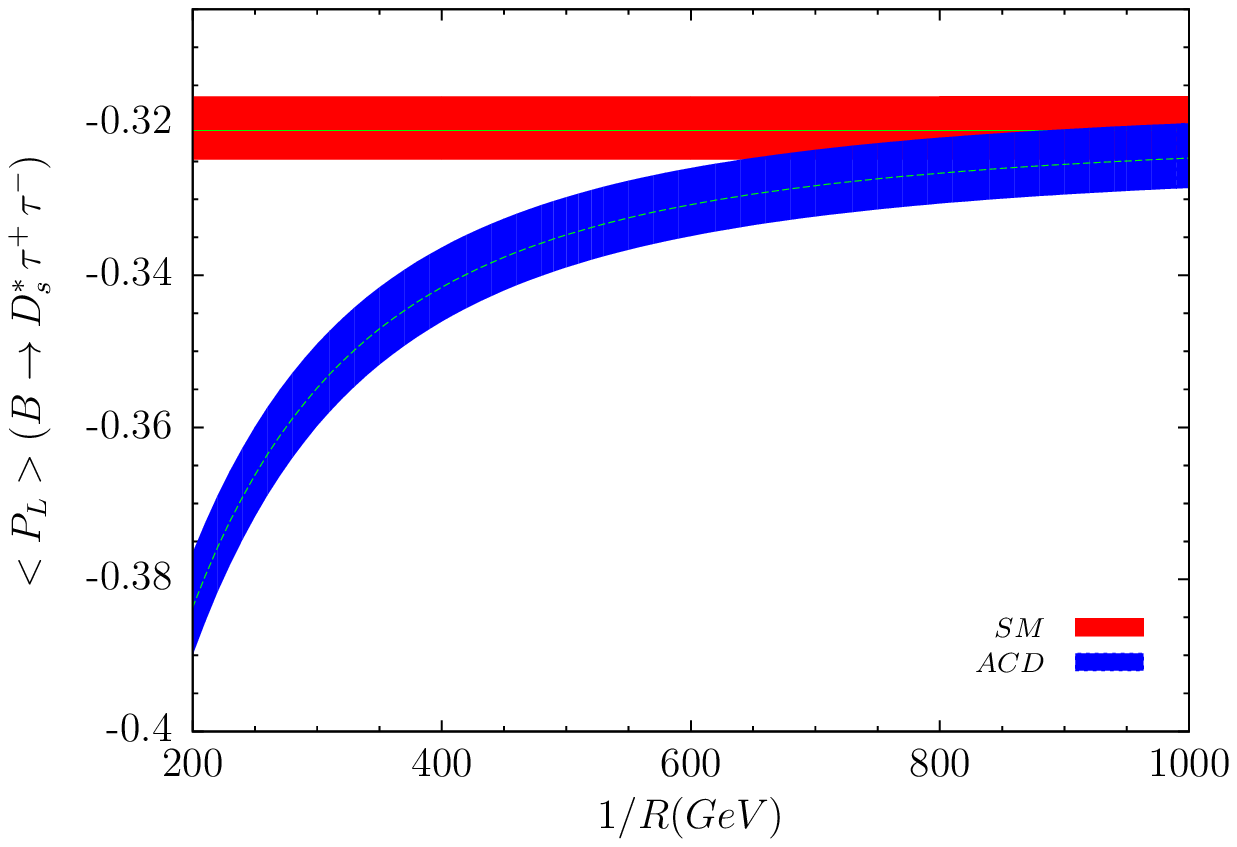}
\caption{(color online) The dependence of longitudinal polarization on $1/R$ including uncertainty on form factors and resonance contributions. \label{LongPol}}
\end{figure}
Similarly, the transversal polarization is given by
\bea
\label{bdtrans}
P^-_T&=&\frac{1}{\Delta_{D^\ast_s}} m_{B_c} m_\ell \pi \sqrt{s \lambda} \Big[
   -8 m^2_{B_c} Re[AB^\ast] + \frac{(1-r-s)}{rs} Re[BF^\ast]-\frac{m^2_{B_c}\lambda}{rs} Re[CF^\ast]\nnb \\
& &-\frac{m^2_{B_c}}{rs}(1-r)(1-r-s) Re[BG^\ast]+\frac{m^4_{B_c}}{rs}\lambda (1-r) Re[CG^\ast] \nnb \\
& &-\frac{m^2_{B_c}}{r}(1-r-s)Re[BH^\ast]+\frac{m^4_{B_c}\lambda}{r} Re[CH^\ast]
\Big]
\eea
\begin{figure}[h]
\centering
\includegraphics[scale=0.62]{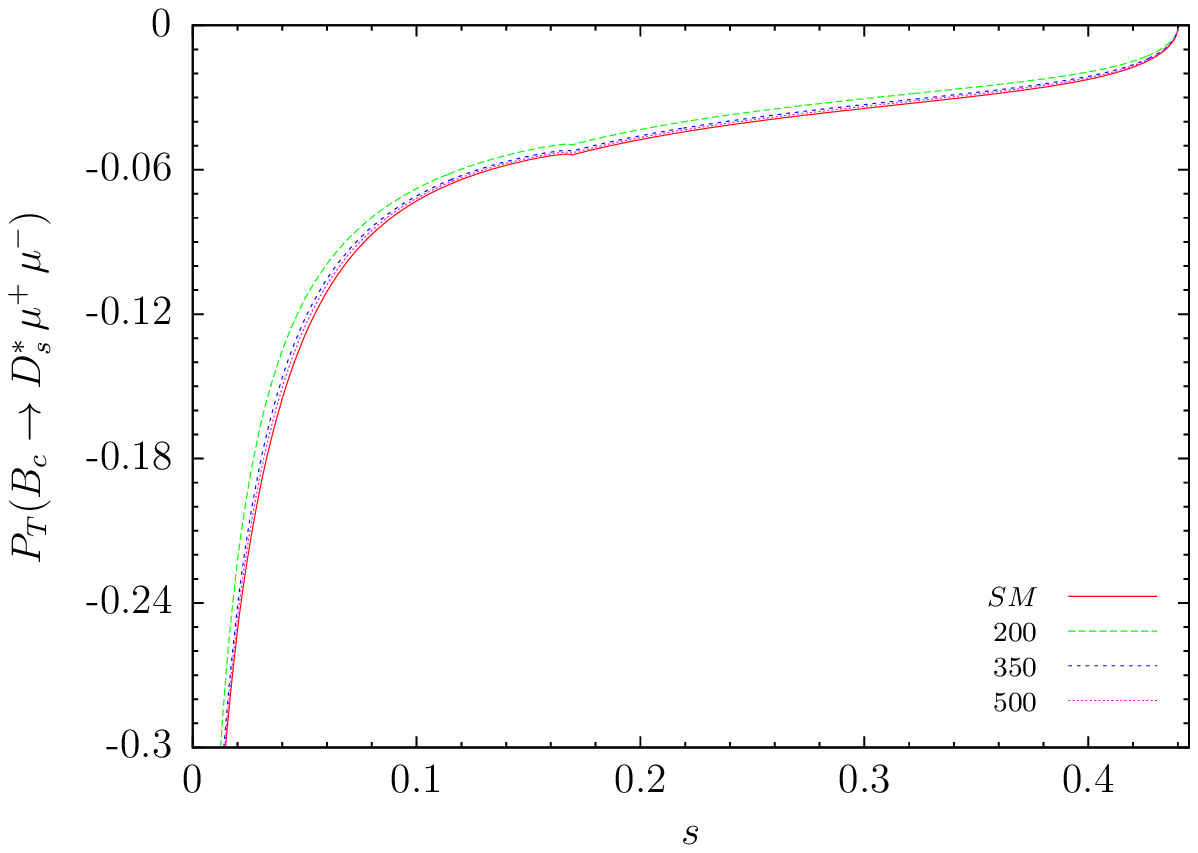}~~~~~
\includegraphics[scale=0.62]{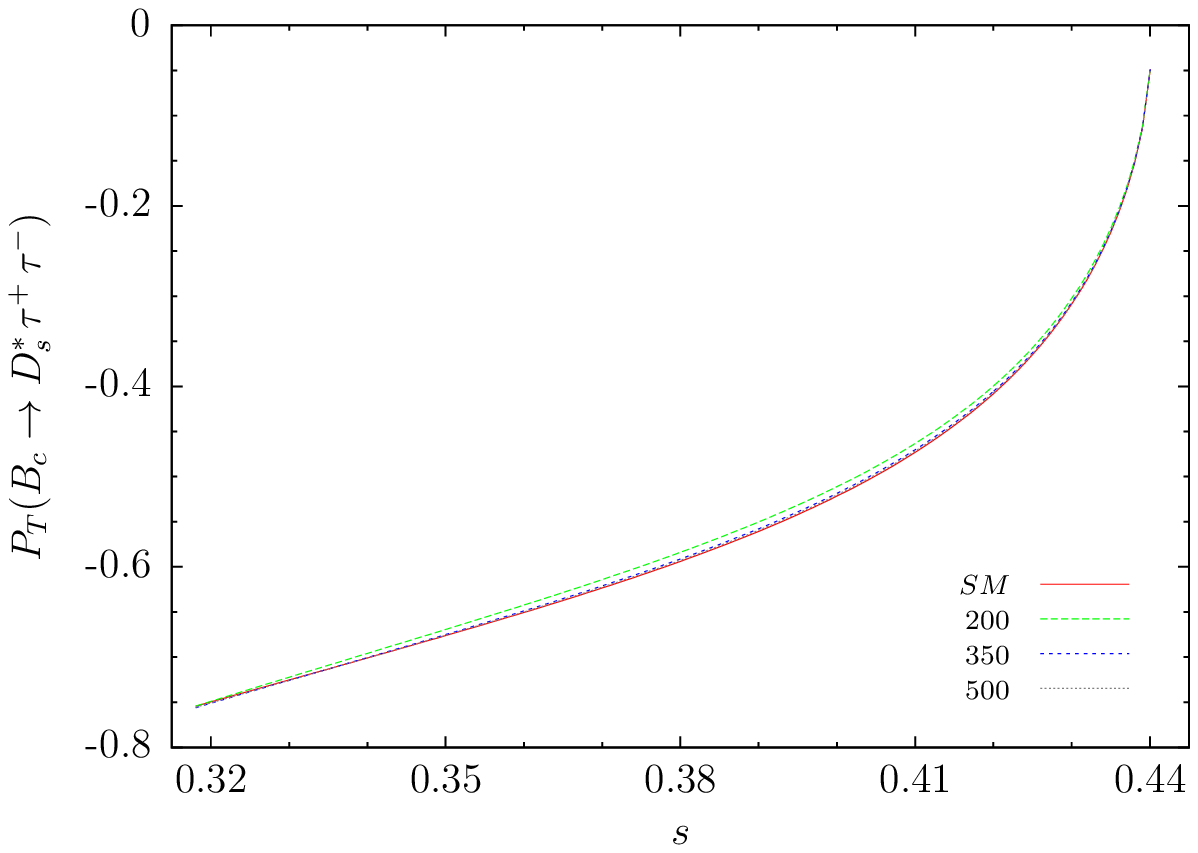}
\caption{(color online) The dependence of transversal polarization on s without resonance contributions using central values of form factors. \label{TransPol-ss}}
\end{figure}
\begin{figure}[h]
\centering
\includegraphics[scale=0.62]{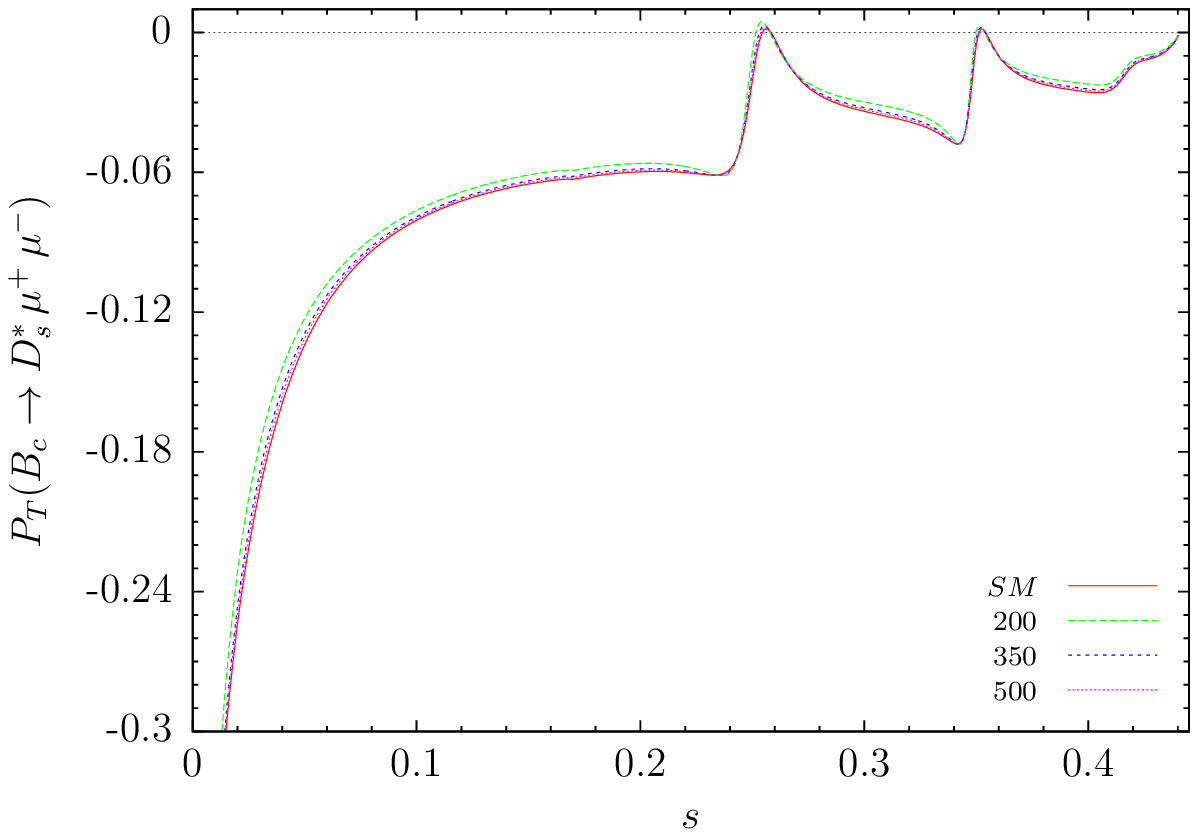}~~~~~
\includegraphics[scale=0.62]{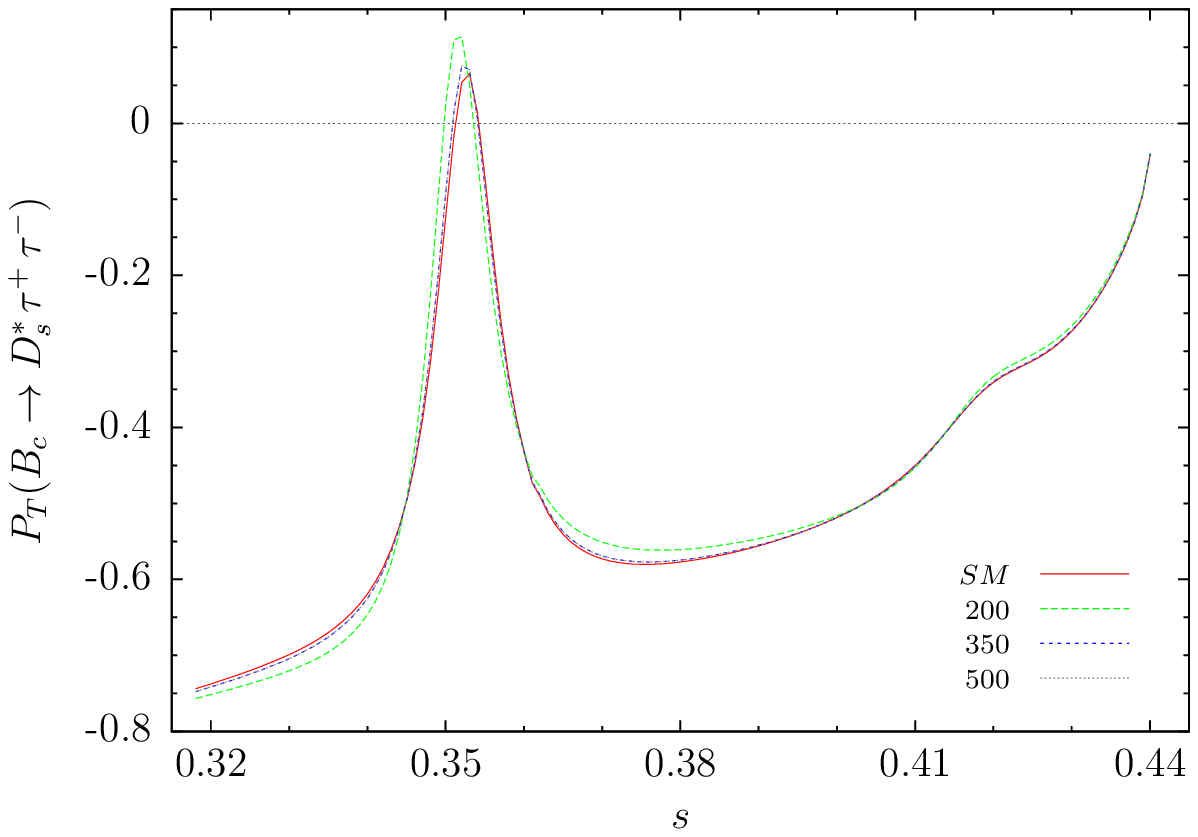}
\caption{(color online) The dependence of transversal polarization on s with resonance contributions using central values of form factors. \label{TransPol-res-ss}}
\end{figure}
\begin{figure}[h]
\centering
\includegraphics[scale=0.62]{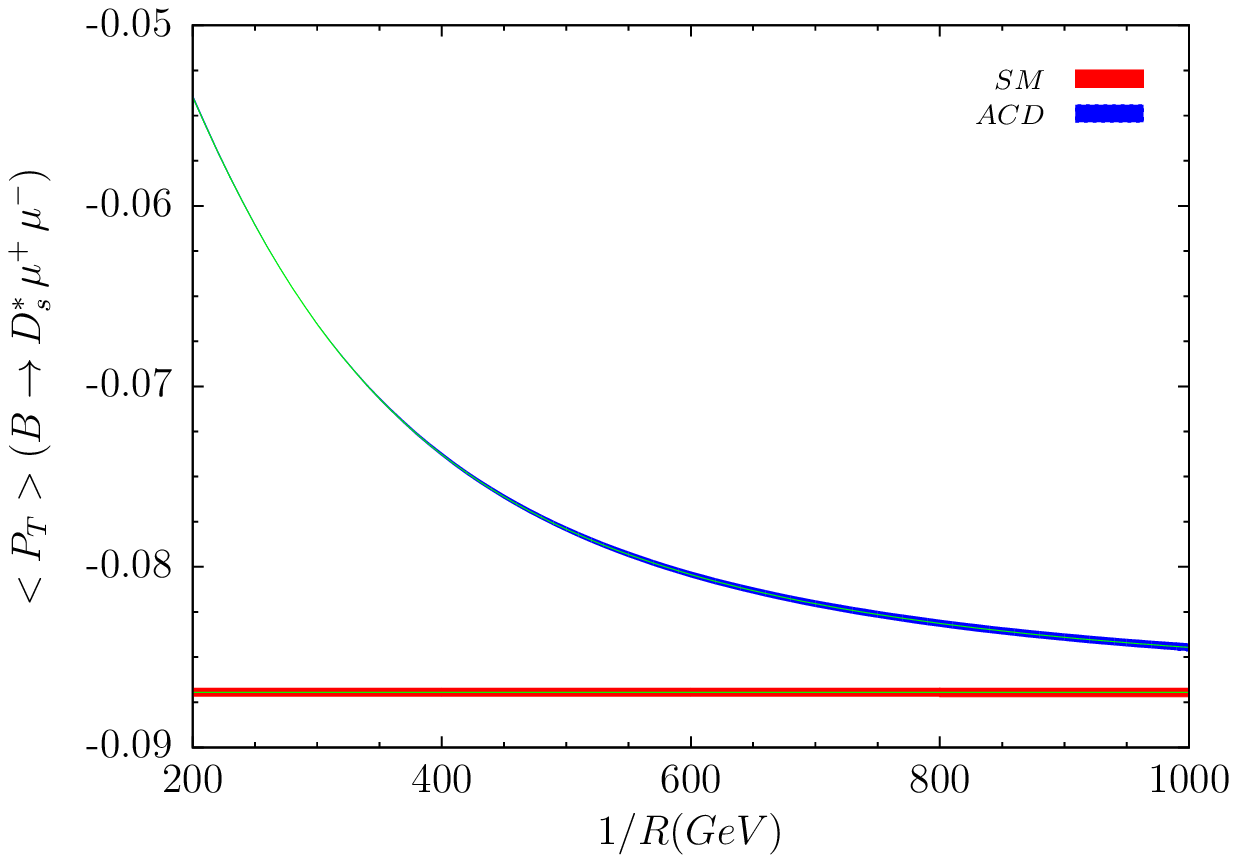}~~~~~
\includegraphics[scale=0.62]{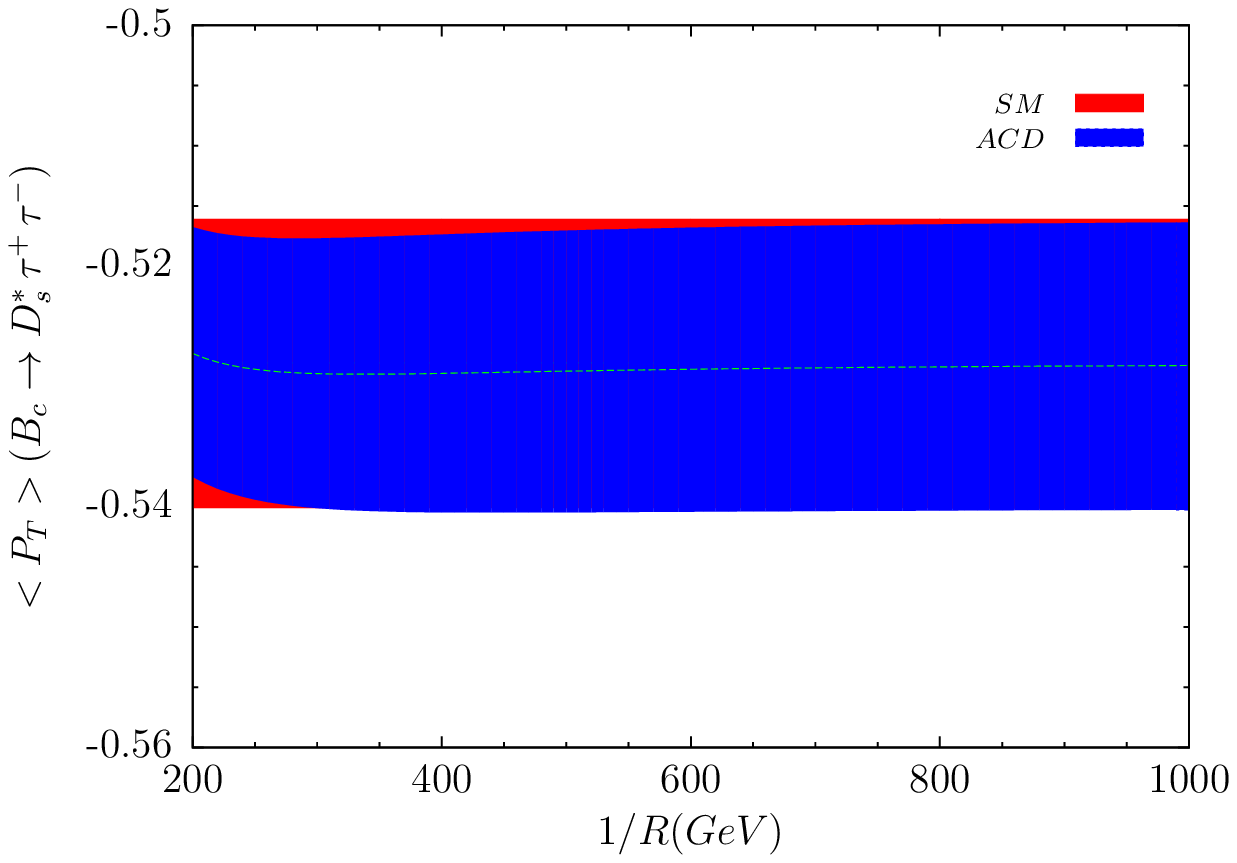}
\caption{(color online) The dependence of transversal polarization on $1/R$, including uncertainty on form factors. \label{TransPol}}
\end{figure}
and the normal polarization by
\bea
\label{bdnorm}
P^-_N&=&\frac{1}{\Delta_{D^\ast_s}} m^3_{B_c} m_{\ell}\pi v \sqrt{s \lambda} \Big[
    -4 Im[B E^\ast]-4 Im[AF^\ast]+\frac{1}{r}(1-r-s) Im[F H^\ast] \nnb \\
&&+\frac{1}{r} (1+3r-s) Im[F G^\ast]-\frac{1}{r}m^2_{B_c} \lambda
        Im[G H^\ast] \Big].
\eea
\begin{figure}[h]
\centering
\includegraphics[scale=0.62]{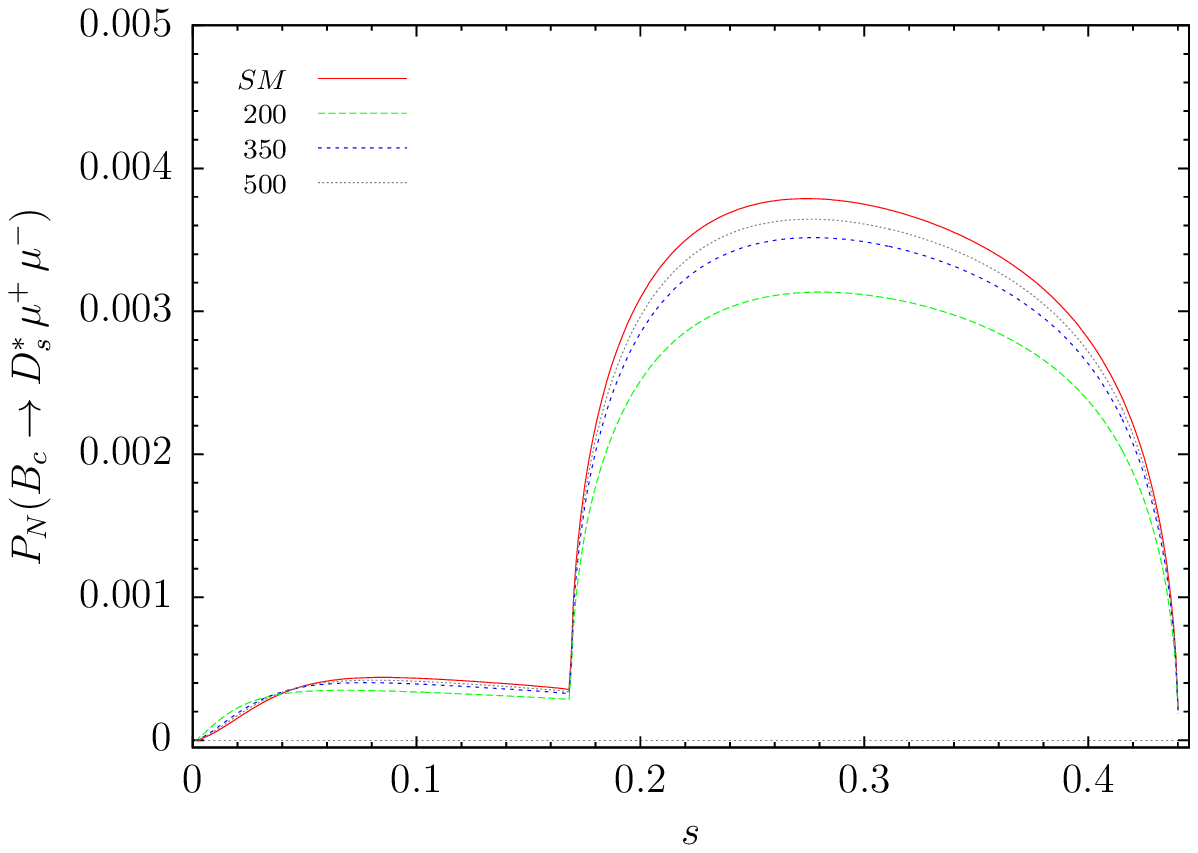}~~~~~
\includegraphics[scale=0.62]{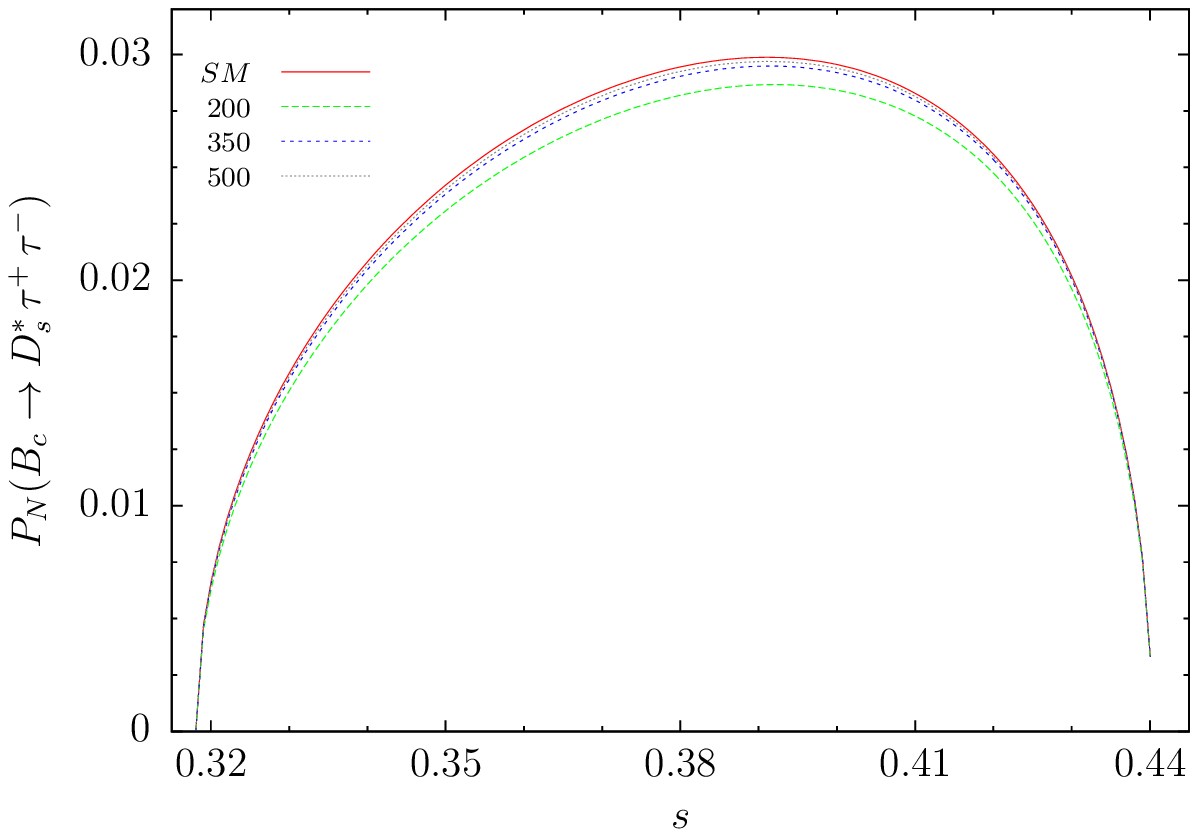}
\caption{(color online) The dependence of normal polarization on s with resonance contributions using central values of form factors. \label{NormPol-ss}}
\end{figure}
\begin{figure}[h]
\centering
\includegraphics[scale=0.62]{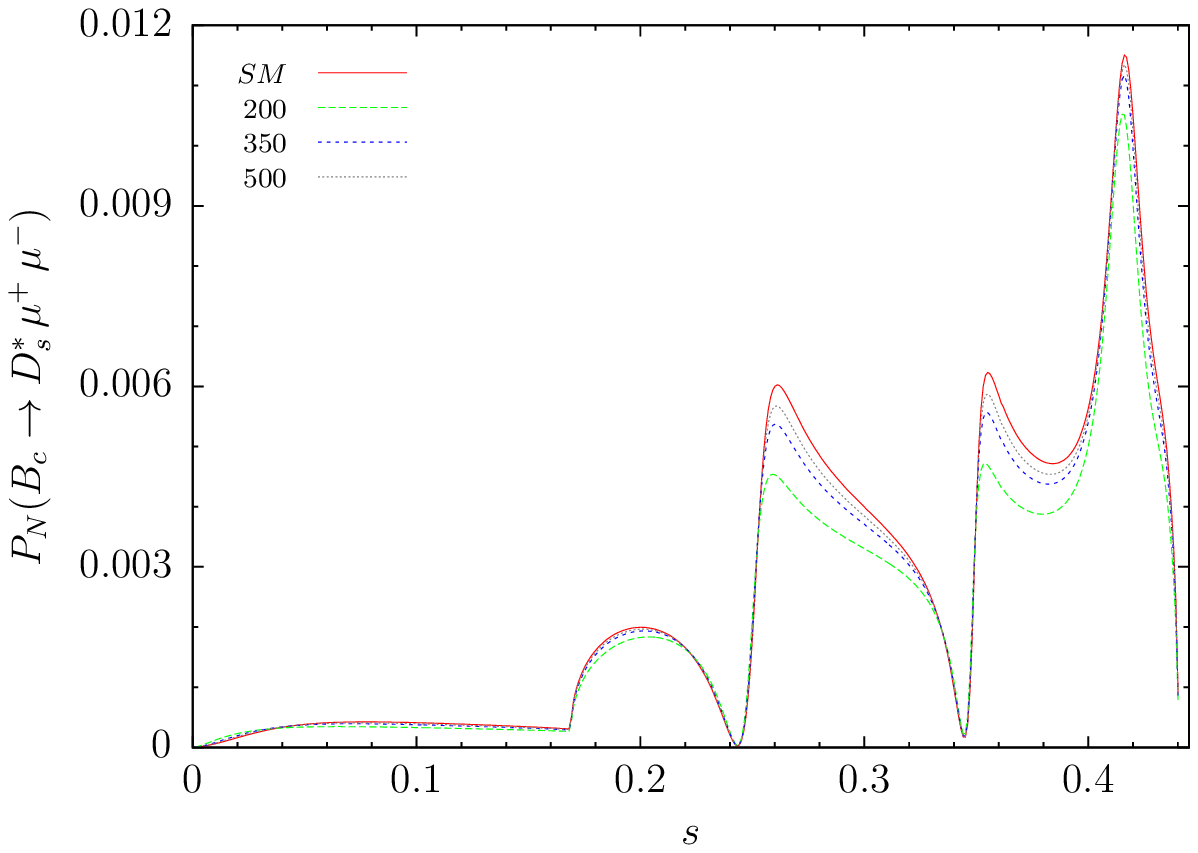}~~~~~
\includegraphics[scale=0.62]{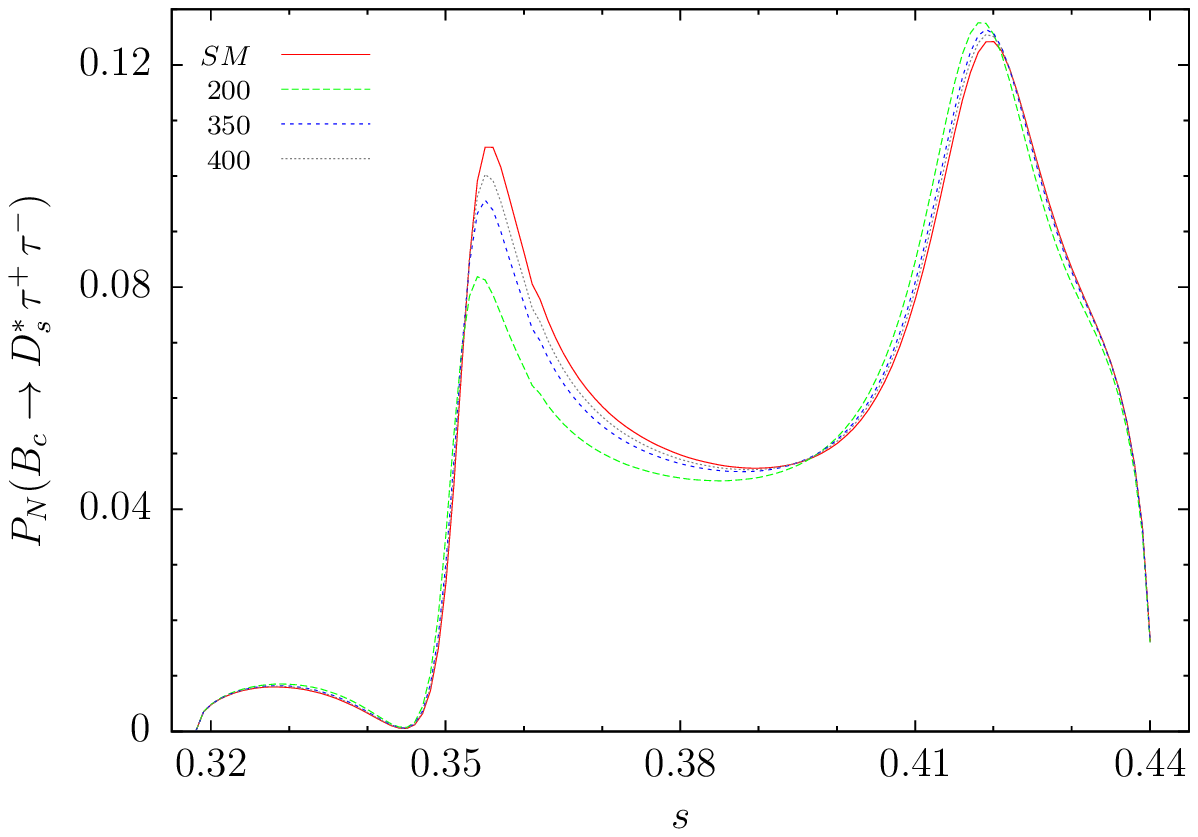}
\caption{(color online) The dependence of normal polarization on s without resonance contributions using central values of form factors. \label{NormPol-res-ss}}
\end{figure}
\begin{figure}[h]
\centering
\includegraphics[scale=0.62]{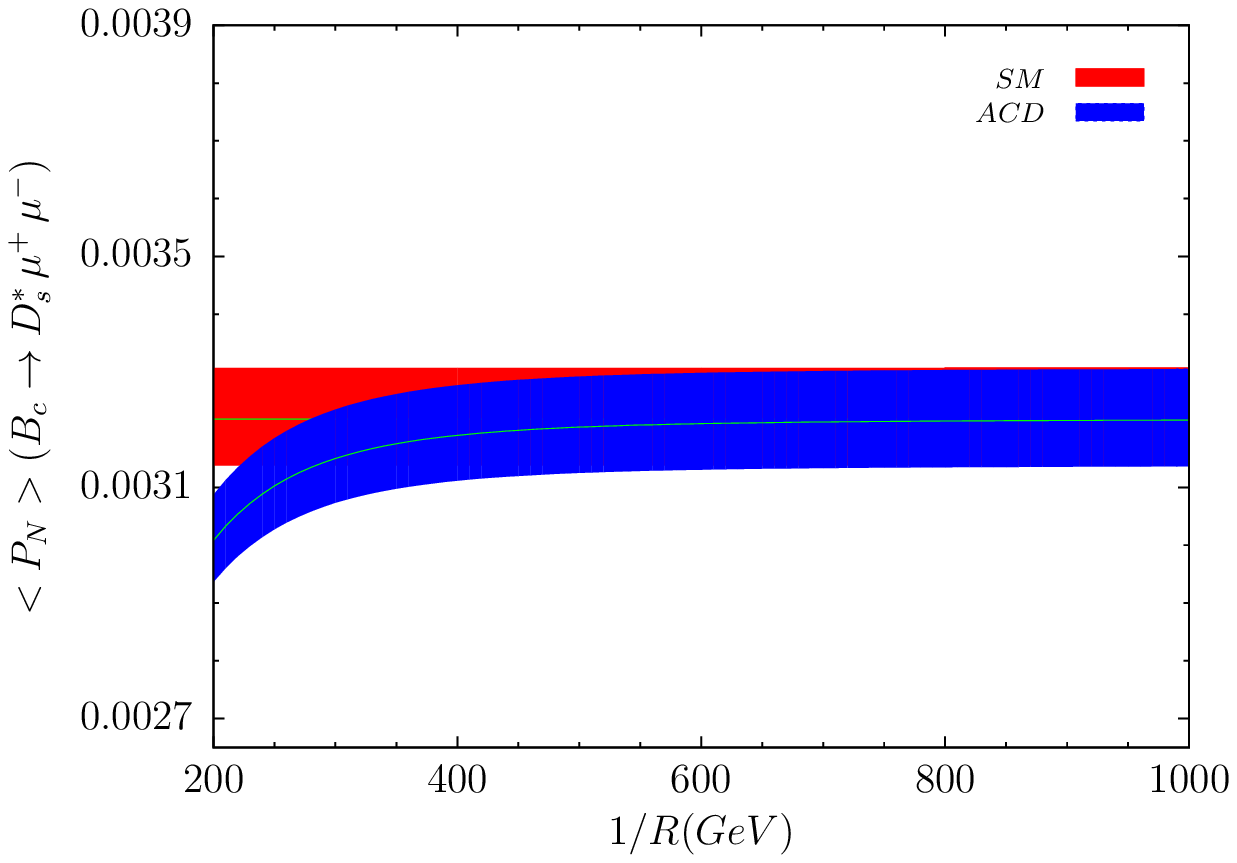}~~~~~
\includegraphics[scale=0.62]{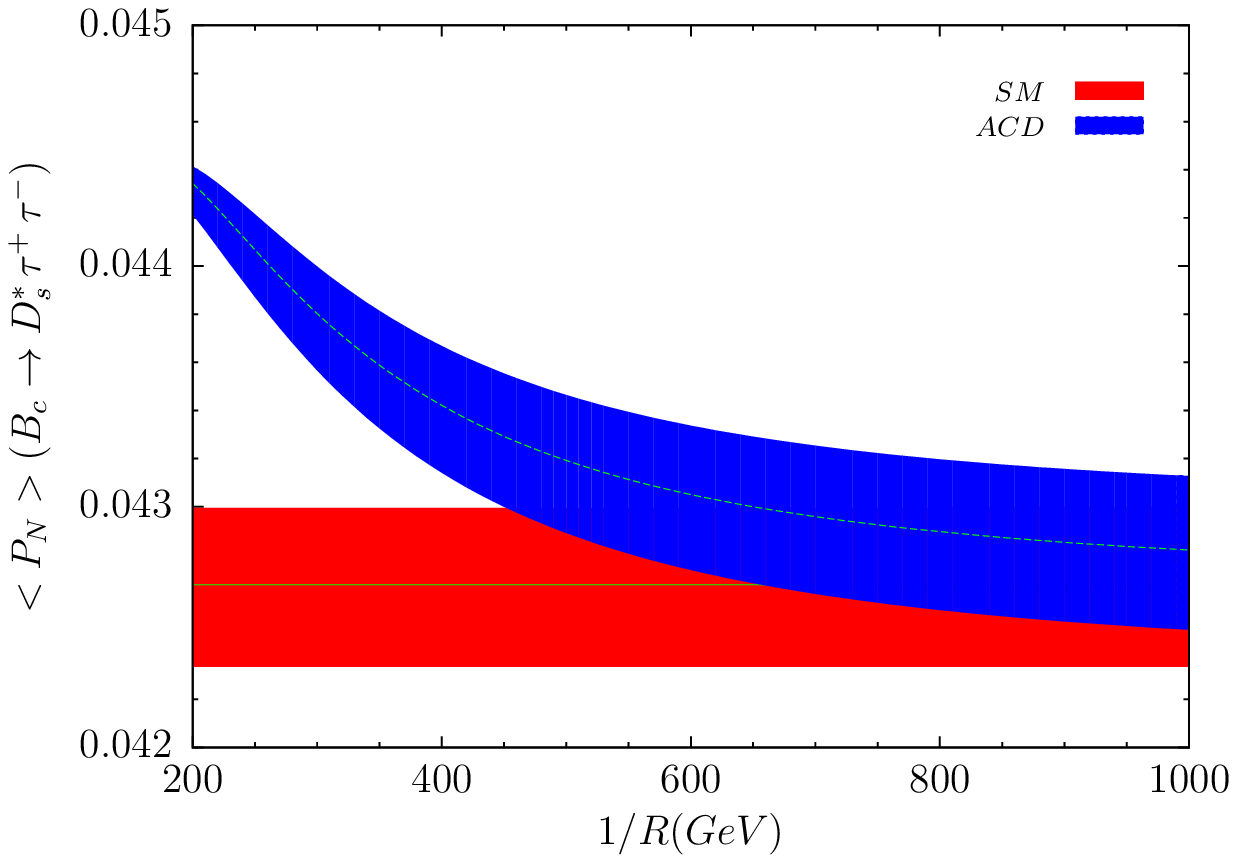}
\caption{(color online) The dependence of normal polarization on $1/R$ including uncertainty on form factors and resonance contributions. \label{NormPol}}
\end{figure}
We eliminate the dependence of the lepton polarizations on $s$ in order to clarify dependence on $1/R$, by considering
the averaged forms over the allowed kinematical region. The averaged lepton polarizations are defined by
\bea
\label{av} \lla P_i \rra =
        \frac{\ds \int_{(2 m_\ell/m_{B_c})^2}^{(1-m_{D_{s}^\ast}/m_{B_c})^2}
            P_i \frac{d{\cal B}}{ds} ds} {\ds \int_{(2
        m_\ell/m_{B_c})^2}^{(1-m_{D_{s}^\ast}/m_{B_c})^2}
 \frac{d{\cal B}}{ds} ds}.
\eea

The dependence of longitudinal polarizations on s with and without resonance contributions are given in Figs. \ref{LongPol-ss} and \ref{LongPol-res-ss}, respectively. For high values of s as $1/R$ approaches 200 GeV the deviation from the SM results get greater for $\tau$ in both resonance and non-resonance cases, while for $\mu$ channel this effect can be seen clearly for all s values when resonance contributions are not added; including resonance contributions, around the peaks this effect seems to be suppressed and only for low values of s we can mention a deviation. Eliminating the dependence of polarization on $s$, we get variation of longitudinal polarization with respect to $1/R$, given by Fig. \ref{LongPol}. For $1/R\geq 500\,GeV$, the difference becomes less important for both channels.
The SM longitudinal polarization, $P_{L}=-0.599$, develops into $-0.670$ ($-0.646$) for $1/R \geq 250 (350)\,GeV$ for $\mu$.
A similar aspect can also be noticed for $\tau$. That is, $P_{L}=-0.321$ SM value vary to $-0.366$ ($-0.347$) for $1/R\geq 250 (350)$.

The dependence of transversal polarization on s with and without resonance contributions are given in Figs. \ref{TransPol-ss} and \ref{TransPol-res-ss}, respectively. The UED effect is unimportant in both decay channels.
In view of $1/R$ dependency, given by Fig. \ref{TransPol}, no difference is observed for $\tau$ decay. Up to $1/R=600\,GeV$ the change is sizeable for $\mu$ channel. In particular, between $1/R=250-350 \,GeV$ the difference might be checked for a signal of new physics.

We have plotted the variation of normal polarizations on s with and without resonance contributions in Figs. \ref{NormPol-ss} and \ref{NormPol-res-ss}, respectively and on $1/R$ in Fig. \ref{NormPol}. The SM value itself for $\mu$ is tiny and as can be seen from the figures the effect of UED on normal polarization in this channel is irrelevant. Additionally, the relatively greater value of normal polarization in the SM for $\tau$ differs slightly.
\section{Conclusion}
In this work, we discussed the \BcDsll decay for $\mu$ and $\tau$ as final state leptons in the SM and the ACD model.
We used form factors calculated in QCD sum rules and throughout the work, we reflected the errors on form factors on calculations and demonstrate the results in possible plotting.

Comparing the SM results and our theoretical predictions on the branching ratio for both decay channels, we obtain the lower bound as $1/R\sim 250\,GeV$. Although this is consistent with the previously mentioned results, a detailed analysis, particularly with the data supplied by experiments, is necessary to put a precise bound on the compactification scale.

As an overall result, we can conclude that, as stated previous works in literature, as $1/R\rar 200\, GeV$ the physical values differ from the SM results.
Up to a few hundreds GeV above the considered bounds, $1/R\geq 250\,GeV$ or $1/R\geq 350)\,GeV$, it is possible to see the effects of UED.

Taking the differential branching ratio into consideration, for small values of $1/R$  there comes out essential difference comparing with the SM results.

Difference between the SM and the ACD results in the forward-backward asymmetry of final state leptons, particularly in the specified region, the obtained result is essential. In addition, the position of the zero of forward-backward asymmetry, which is sensitive in searching new physics, can be a useful tool to check the UED contributions.

Polarization of the leptons have been studied comprehensively and we found that transversal and normal polarizations are not sensitive to the extra dimension, only dependence of transversal (normal) polarization on $1/R$ for $\mu$ ($\tau$) decay channel for low values of $1/R$ might be useful. However, studying longitudinal polarization for both leptons up to $1/R=600\,GeV$ will be a powerful tool establishing new physics effects.

Under the discussion throughout this work, the sizable discrepancies between the ACD model and the SM predictions
at lower values of the compactification scale can be considered the indications of new physics and should be searched in the experiments.
\\
\begin{acknowledgments}
The author would like to thank K. Azizi for valuable discussion and
U. Kanbur for contributions on computer base works.
\end{acknowledgments}
\clearpage
\newpage
%
\appendix
\section {Wilson Coefficients in the ACD Model}

In the ACD model, the new physics contributions appear by modifying available Wilson coefficients in the SM. The modified Wilson coefficients are calculated in \cite{Buras03}-\cite{Buras04} and can be expressed in terms of $F(x_t, 1/R)$ which generalize the corresponding SM functions $F_0(x_t)$ according to
\bea
\label{WACD-app} F(x_t, 1/R) = F_0(x_t) + \sum_{n=1} ^{\infty} F_n(x_t, x_n)
\eea
where $x_t=m_t^2/m_W^2$, $x_n=m_n^2/m_W^2$ and $m_n=n/R$.

Instead of $C_7$, an effective, normalization scheme independent, coefficient $C^{eff}_7 $ in the leading logarithmic approximation is defined as
\bea
\label{C7eff} C_7^{eff}(\mu_b, 1/R) = && \eta^{16/23} C_7(\mu_W, 1/R) \nnb \\
        && +   \frac{8}{3} (\eta^{14/23} - \eta^{16/23}) C_8(\mu_W,1/R) +
                    C_2(\mu_W, 1/R) \sum_{i=1} ^{8} h_i \eta^{a_i}
\eea
with $\eta =\frac {\alpha_s(\mu_W)} {\alpha_s(\mu_b)}$ and
\bea
\label{alphas} \alpha_s(x)=\frac {\alpha_s(m_Z)} {1-\beta_0 \frac{\alpha_s(m_Z)} {2 \pi} ln(\frac{m_Z} {x})}
\eea
where in fifth dimension $\alpha_s(m_Z)=0.118$ and $\beta_0=23/3$.

The coefficients $a_i$ and $h_i$ are  \\
\bea
a_i   & = &\Big(\frac{14}{23},\frac{16}{23},\frac{6}{23},-\frac{12}{23},0.4086,-0.4230,-0.8994,0.1456 \Big) \nnb \\
h_{i} & = & \Big(2.2996,-1.088,-\frac{3}{7},-\frac{1}{14},-0.6494,-0.0380,-0.0186,-0.0057 \Big).
\eea
The functions in (\ref{C7eff}) are
\bea
\label{C2mw} C_2 (\mu_W) = 1,~~~
C_7(\mu_W, 1/R) = - \frac{1} {2} D^\prime (x_t, 1/R),~~~
C_8(\mu_W, 1/R) = - \frac{1} {2} E^\prime (x_t, 1/R).
\eea
Here, $D^\prime (x_t, 1/R)$ and $E^\prime (x_t, 1/R)$ are defined by using (\ref{WACD-app}) with the following functions
\bea
\label{Dpzero} D_0^{\prime}(x_t) = - \frac {(8x_t^3 + 5x_t^2 - 7x_t)} { 12(1-x_t)^3} +
                    \frac {x_t^2 (2-3x_t)} {2(1-x_t)^4} lnx_t
\eea
\bea
\label{Epzero} E_0^{\prime} (x_t) = - \frac{x_t (x_t^2 - 5x_t - 2)} {4(1-x_t)^3} +
                    \frac {3 x_t^2} {2(1-x_t)^4} lnx_t
\eea
\bea
\label{Dpn} D_n^{\prime} (x_t, x_n) = \frac{ x_t (-37 + 44x_t + 17x_t^2 + 6x_n^2(10-9x_t+3x_t^2) - 3x_n(21-54x_t+17x_t^2))} {36(x_t - 1)^3} \nnb \\
-\,  \frac{(-2+x_n+3x_t)(x_t+3x_t^2 + x_n^2(3+x_t) - x_n(1+ (-10+x_t)x_t))} {6(x_t-1)^2} \ln \frac{x_n + x_t} {1+x_n}\nnb \\
+\, \frac{x_n (2-7x_n+3x_n^2} {6} \ln \frac{x_n} {1+x_n}~~~~~~~~~~~~~~~~~~~~~~~~~~~~~~~~~~~~~~~~~~~~~~~~~~~~~~~~~~~~
\eea
\bea
\label{Epn} E_n^{\prime} (x_t, x_n)= \frac{x_t(-17-8x_t+x_t^2-3x_n(21-6x_t+x_t^2) - 6x_n^2(10-9x_t+3x_t^2))} {12(x_t-1)^3} ~~~~~~~~~\nnb \\
+\, \frac{(1+x_n)(x_t+3x_t^2+x_n^2(3+x_t)-x_n(1+(-10+x_t)x_t))}{2(x_t-1)^4} \ln \frac{x_n+x_t} {1+x_n} \nnb \\
-\, \frac{1}{2}x_n(1+x_n)(-1+3x_n) \ln \frac{x_n}{1+x_n}.~~~~~~~~~~~~~~~~~~~~~~~~~~~~~~~~~~~~~~~~~~~
\eea
Following \cite{Buras03} or directly from \cite{Colangelo06} one gets the expressions for the sum over n as
\bea \label{Dpsum} \sum _{n=1} ^{\infty} D_n^{\prime}(x_t, x_n)&&= -\frac{x_t(-37+x_t(44+17x_t))}{72(x_t-1)^3} \nnb \\
   &&+\frac{\pi M_W R}{2} \Bigg[ \int_0^1 dy \,\frac{(2 y^{1/2} + 7 y^{3/2} + 3 y^{5/2})}{6} \coth (\pi M_W R \sqrt y)\nnb \\
    &&+\, \frac{(-2+3x_t)x_t(1+3x_t)}{6(x_t-1)^4} J(R, -1/2)\nnb \\
      &&  -\, \frac{1}{6(x_t-1)^4} [x_t (1+3x_t)-(-2+3x_t)(1+(-10+x_t)x_t)]J(R, 1/2) \nnb \\
        &&    +\, \frac{1} {6(x_t-1)^4}[(-2+3x_t)(3+x_t) - (1+(-10+x_t)x_t)]J(R, 3/2) \nnb \\
          &&      -\, \frac{(3+x_t)}{6(x_t-1)^4} J(R, 5/2)\Bigg]
\eea
and
\bea
\label{Epsum} \sum _{n=1} ^{\infty} E_n^{\prime}(x_t, x_n) &&= -\frac{x_t(-17+(-8+x_t)x_t)}{24(x_t-1)^3} \nnb \\
   &&+\frac{\pi M_W R}{4} \Bigg[ \int_0^1 dy \,( y^{1/2} + 2 y^{3/2} - 3 y^{5/2}) \coth (\pi M_W R \sqrt y)\nnb \\
    &&-\, \frac{x_t(1+3x_t)}{(x_t-1)^4} J(R, -1/2)\nnb \\
      &&  +\, \frac{1}{(x_t-1)^4} [x_t (1+3x_t)-(1+(-10+x_t)x_t)]J(R, 1/2) \nnb \\
        &&    -\, \frac{1} {(x_t-1)^4}[(3+x_t) - (1+(-10+x_t)x_t)]J(R, 3/2) \nnb \\
          &&      +\, \frac{(3+x_t)}{(x_t-1)^4} J(R, 5/2)\Bigg]
\eea
where
\bea \label{ijr} J(R, \alpha) = \int_{0}^{1} dy \,y^{\alpha} \,[\coth (\pi M_W R \sqrt y) - x_t^{1+\alpha} \coth (\pi m_t R \sqrt y)].
\eea
The Wilson coefficient $C_9$ in the ACD model and the NDR scheme is
\bea
\label{C9mu} C_9(\mu, 1/R)= P_0^{NDR} + \frac {Y(x_t, 1/R)} {sin^2{\theta_W}} - 4 Z(x_t, 1/R) + P_E E(x_t, 1/R)
\eea
where $P_0^{NDR}=2.6 \pm 0.25$ and $P_E$ is numerically negligible. The functions $Y(x_t, 1/R)$ and $Z(x_t, 1/R)$ are defined as
\bea
\label{Yxtr} Y(x_t, 1/R) = Y_0(x_t) + \sum _{n=1} ^{\infty} C_n(x_t, x_n)
\eea
\bea \label{Zxtr} Z(x_t, 1/R) = Z_0(x_t) + \sum _{n=1} ^{\infty} C_n(x_t, x_n)
\eea
with
\bea \label{Yzero} Y_0(x_t) = \frac {x_t} {8} \Bigg[ \frac{x_t - 4} {x_t - 1} +
                    \frac {3 x_t} {(x_t - 1)^2} lnx_t \Bigg]
\eea
\bea \label{Zzero} Z_0 (x_t) = \frac {18x_t^4 - 163x_t^3 + 259x_t^2 - 108x_t} {144(x_t - 1)^3} +
                    \Bigg[ \frac {32x_t^4 - 38x_t^3 - 15x_t^2 + 18x_t} {72(x_t - 1)^4} - \frac{1} {9} \Bigg]lnx_t
\eea
\bea \label{Cn} C_n(x_t, x_n) = \frac{x_t} {8(x_t-1)^2} \Bigg[x_t^2 - 8x_t +
                                7 + (3 + 3x_t + 7x_n - x_t x_n)ln\frac{x_t + x_n} {1+x_n}\Bigg]
\eea
and
\bea \label{Csum}\sum _{n=1} ^{\infty} C_n(x_t, x_n) = \frac{x_t (7-x_t)} {16(x_t-1)}
     - \frac{\pi M_W R x_t} {16(x_t-1)^2} [3(1+x_t) J(R, -1/2) + (x_t-7) J(R, 1/2)].
\eea
The $\mu$ independent $C_{10}$ is given by
\bea
\label{C10} C_{10}(1/R) = - \frac{Y(x_t, 1/R)} {sin^2{\theta_W}}
\eea
where $Y(x_t, 1/R)$ is defined in (\ref{Yxtr}).

\end{document}